\documentclass[aps,twocolumn,showpacs,final,floatfix]{revtex4-1}
\usepackage[pdftex]{graphicx}  
\usepackage[update]{epstopdf}
\usepackage{amssymb}
\usepackage{amsmath}
\usepackage{color}
\usepackage{textcomp}
\usepackage[utf8]{inputenc}
\newcommand\ra{\rangle}
\newcommand\la{\langle}
\newcommand\nn{\nonumber}

\graphicspath {{./figures/}}
\makeatletter
\def\input@path{{./figures/}}
\makeatother

\begin{document}
 \title{ Transport, correlations, and chaos in a classical disordered anharmonic chain }

\author{Manoj Kumar$^{1,2}$, Anupam Kundu$^1$, Manas Kulkarni$^1$, David A. Huse$^{3,4}$ and Abhishek Dhar$^1$}

\affiliation{$^1$International Centre for Theoretical Sciences, Tata Institute of Fundamental Research, Bengaluru -- 560089, India.
\\$^2$Centre for Fluid and Complex Systems, Coventry University, CV1 5FB, United Kingdom.\\$^3$Physics Department, Princeton University, Princeton, NJ 08544, USA,\\$^4$Institute for Advanced Study, Princeton, NJ 08540, USA}
\date{\today}
\begin{abstract}
We explore transport properties in a disordered nonlinear chain of classical harmonic oscillators, and thereby identify a regime exhibiting  behavior analogous to that seen in quantum  many-body-localized systems.
Through extensive numerical simulations of this system connected at its ends to heat baths at different temperatures, we computed the  heat current and the temperature profile in the nonequilibrium steady state as a function of system size $N$, disorder strength $\Delta$, and temperature $T$.  The conductivity $\kappa_N$, obtained for finite length ($N$) systems, saturates to a value $\kappa_\infty >0$ in the large $N$ limit, for all values of disorder strength $\Delta$ and temperature $T>0$. We show evidence that for any $\Delta>0$ the conductivity goes to zero faster than any power of $T$ in the $(T/\Delta) \to 0$ limit, and find that the form $\kappa_\infty \sim e^{-B |\ln(C \Delta/T)|^3}$ fits our data. This form has earlier been suggested by a theory based on the dynamics of multi-oscillator chaotic islands. 
The finite-size effect can be $\kappa_N < \kappa_{\infty}$ due to boundary resistance when the bulk conductivity is high (the weak disorder case), or $\kappa_N > \kappa_{\infty}$ due to direct bath-to-bath coupling through bulk localized modes when the bulk is weakly conducting (the strong disorder case).
We also present results on equilibrium dynamical correlation functions and on the  role of chaos on transport properties. Finally, we explore the differences in the growth and propagation of chaos in the weak and strong chaos regimes by studying the classical version of the Out-of-Time-Ordered-Commutator. 
\end{abstract}

\maketitle

 \section{Introduction}
 \label{s1}

In the last two decades,   there has been a considerable amount of interest in understanding the transport properties of systems  in the presence of both disorder and interactions.  It is well-known that disordered systems described by quadratic Hamiltonians (e.g noninteracting electrons in a disordered potential or disordered harmonic crystals) exhibit the phenomena of Anderson localization \cite{anderson1958absence}, whereby the single-particle eigenstates or normal modes (NMs) of the system form spatially localized states.  This has a profound effect on transport --- in particular for one dimensional systems all states are localized and one finds that the system is a thermal insulator. 

A question of great interest is to ask what happens when one introduces interactions in such a system: Does one need a nonzero critical strength of interactions to see an insulator-to-conductor transition? For quantum systems, this question has been extensively studied in the context of many-body localization (MBL) \cite{nandkishore2015many,alet2018many}. It is now generally accepted that for one-dimensional quantum systems with a sufficiently strong disorder, the localized insulating state persists up to a critical interaction strength. One can ask the same question in the context of classical systems and this has been addressed some recent works \cite{PRL.100.134301,Huse2009,flach2011thermal,Huveneers2013,Narayan2018}.
The work in \cite{PRL.100.134301,Huse2009,flach2011thermal,Huveneers2013} 
leads one to believe that there is no classical analogue of an MBL phase, while \cite{Narayan2018} provides evidence that such a phase might exist in a nonlinear oscillator chain, for a specially-designed realization of spring constants.  Theoretical arguments in \cite{Huveneers2013} indicate that the thermal conductivity of a disordered nonlinear system goes to zero with decreasing temperature $T$ faster than any power of $T$.  The numerical study in \cite{Huse2009} is consistent with this finding, however Flach et al. in \cite{flach2011thermal} found evidence for a power-law dependence. A recent study proved sub-diffusive transport in a disordered chain with sparse interacting regions \cite{de2020subdiffusion}.

Several other numerical as well as theoretical studies have also investigated  the phenomena of Anderson  localization, wave-packet diffusion, and transport in  nonlinear disordered media  \cite{PRL.100.134301,Huse2009,flach2011thermal,Huveneers2013,PRL.64.1397,PRL.100.134301,PRL.100.094101,PRE.47.4108,PRE.79.056211,PRL.64.1693,devillard1986polynomially,doucot1987anderson,knapp1991transmission,PRL.69.1807,laptyeva2012subdiffusion,flach2009universal}.  Numerical studies have shown that nonlinearity gives rise to the subdiffusive spreading of a wave packet in an otherwise empty lattice  (thus zero temperature), implying the destruction of Anderson localization \cite{PRL.100.094101,PRE.79.056211,flach2009universal}.  A theoretical explanation of the subdiffusive spreading is based on the fact that the nonlinearity results in non-integrability of the system  because of which the wave packet evolves chaotically, and this leads to an incoherent spreading \cite{frohlich1986localization,PRL.100.084103,flach2009universal,flach2010spreading,tietsche2008chaotic,flach2010,Basko2011,Basko2012}.  A possible mechanism of chaos generation and thermalization at nonzero temperature was discussed in \cite{Basko2011}, in the context of the disordered discrete nonlinear Schrodinger equation.  Based on this picture it was estimated  that the probability of chaos generation scales  at a low temperature as $e^{-B |\ln(C \Delta/T)|^3} $,  where $B,C$ are constants, $\Delta$ is disorder, and $T$ denotes the temperature.  It is, therefore, argued that the conductivity follows the same scaling.

 The main aim of this paper is to  extract,  through extensive numerics, the temperature dependence of the thermal conductivity of the disordered anharmonic chain in the   $T \to 0$ limit, and to understand the precise mechanism of transport in this system. 
 We also aim to look for signatures of MBL during a crossover from strong to weak chaos  at finite temperatures  as the disorder $\Delta$ is varied across a characteristic value $\Delta_c$. 
 For our study, we consider a one-dimensional chain of harmonic oscillators with random frequencies and purely anharmonic nearest-neighbor coupling.   This model lies  in the class introduced by Fr\"{o}hlich, Spencer, and Wayne \cite{frohlich1986localization}, and therefore we referred it as the FSW model. It is the strong disorder limit of the so-called Klein-Gordon (KG) model ~\cite{flach2010,flach2011thermal,laptyeva2014nonlinear,bodyfelt2011nonlinear}. At zero temperature, the model effectively consists of disconnected oscillators at random frequencies and hence a small localization length at low temperatures.  Thus the FSW model is suitable to study the low-temperature behavior of conductivity  since we expect that relatively small system sizes can be used to obtain the asymptotic (infinite size) conductivity. 

We have performed extensive nonequilibrium simulations for a range of temperatures $T$ and the disorder strength $\Delta$, and for different system sizes $N$. We show that finite-size effects in both the low and high disorder regimes can be understood as arising from boundary effects, and use finite-size scaling to extract the  thermal conductivity in the infinite size limit. 
As one of our main results we find that our data at the lowest temperature fits well  to the form $ \kappa_\infty = A e^{-B |\ln(C \Delta/T)|^3}$, which   has earlier been argued  on the basis of the dynamics of multi-oscillator chaotic resonances \cite{Basko2011}.
 In addition to the nonequilibrium simulations, we have also examined the form of  equilibrium dynamical correlation functions for the weak and strong disorder, and our crucial observation is that the behavior of dynamical correlations is truly Gaussian for a weak disorder, while
for the strong disorder, the behavior becomes non-Gaussian but
has a diffusive scaling. Finally, we have looked at the spatio-temporal  propagation of chaos in the system by computing a classical analogue of  Out-of-Time-Ordered-Commutator (OTOC) for our nonlinear disordered model.

The contents of the paper are as follows. In Sec.~\ref{def}, we describe the Hamiltonian and the reservoirs (which are modelled as Langevin baths). We also introduce important dimensionless units which transparently shows that temperature is equivalent to nonlinearity strength. In Sec.~\ref{sec:sim}, we present simulation results for the nonequilibrium steady state heat current. We analyze various aspects such as system size scaling, disorder, and  temperature dependence of the thermal conductivity. In this section, we also present results for the temperature profiles in the nonequilibrium steady state. Sec.~\ref{sec:corr} is devoted to the energy correlations in space and time and, in particular, their dependence on the strength of the disorder. In Sec.~\ref{sec:otoc}, a classical analogue of  Out-of-Time-Ordered-Commutator (OTOC) is investigated for our nonlinear disordered system. In particular, the behavior of the heat map, butterfly velocity, and Lyapunov exponents as one changes disorder strength are analyzed. Finally, in Sec.~\ref{conclu}, we conclude this paper with a brief discussion on our main findings.

\begin{figure}
  \centering
  \includegraphics[width=0.9\columnwidth]{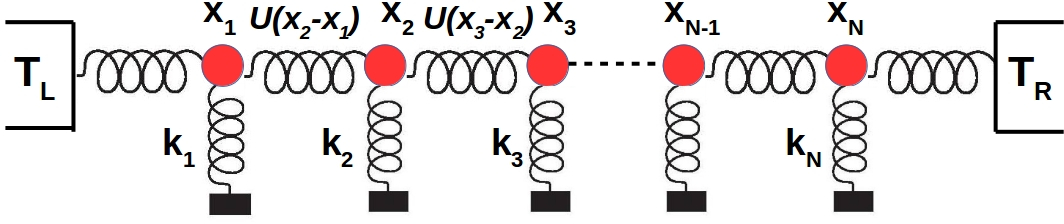}
  \caption{Schematic diagram of the model. $U$ is the nonlinear interactions between oscillators given by $U(x_l-x_{l-1})=(x_l-x_{l-1})^4/4$.}
  \label{fig1}
\end{figure}

\section{Definition of the FSW Model and nonequilibrium dynamics} 
\label{def}

We start by taking a chain of  $N$  oscillators with masses $m$ and random spring constants $k_i=m \omega^2 \varepsilon_i$, with each $\varepsilon_i$ chosen uniformly in the interval $[1-\Delta,1+\Delta]$, where $\Delta$ defines the disorder strength.  Nearest-neighbor oscillators are then coupled by a nonlinear (quartic) interaction potential $U$ of strength $\nu$ (see Fig.~\ref{fig1}). The Hamiltonian of the system is given by 
\begin{equation} 
 \mathcal{H} = \sum_{i=1}^N\left[\frac{p_i^2}{2m}+k_i\frac{x_i^2}{2}\right]+\sum_{i=0}^{N}\nu \frac{(x_i-x_{i+1})^4}{4}, 
 \label{fsw}
\end{equation}
where $\{x_i,p_i\}$ are respectively the positions and momenta of the oscillators in the chain and we set $x_0=x_{N+1}=0$. The limit $\Delta=0$ represents the pure case and  $\Delta=1$ is the maximum disorder strength for this disorder distribution.

The chain of oscillators is attached to two thermal reservoirs at unequal temperatures $T_L$ and $T_R$ at the left and right ends, respectively, so that a temperature gradient is generated, and  there is a heat current along the chain \cite{lepri2003thermal,dhar2008heat}. The two thermal reservoirs are modeled by Langevin equations. In dimensionless units, $t \to \omega t$ and $x \to \sqrt{\nu/(m \omega^2)} x$, the equations of motion for $1 \leq i \leq N$ are  given by
\begin{equation}
 \ddot{x_i}=-\varepsilon_i x_i-[(x_i-x_{i-1})^3+(x_i-x_{i+1})^3]-\gamma_i\dot{x_i}+\eta_i,
 \label{dynamic_eq}
\end{equation}
with $\eta_i=\eta_L\delta_{i,1}+\eta_R\delta_{i,N}$ and $\gamma_i=\gamma(\delta_{i,1}+\delta_{i,N})$. The Gaussian white noise, $\eta_{L,R}$, satisfies the fluctuation-dissipation relation $\langle \eta_{L,R}(t)\eta_{L,R}(t')\rangle=2\gamma T_{L,R}\delta(t-t')$ with $\langle \eta_{L,R} \rangle =0$. Here the dissipation constant $\gamma$ is measured in units of $m\omega$ and temperature in units of $m^2 \omega^4/(\nu k_B)$, where $k_B$ is the Boltzmann constant.  The only relevant dimensionless parameters in the problem that remain with this scaling are the disorder strength $\Delta$, the temperature $T$ (which is equivalent to the nonlinearity strength $\nu$),  dissipation constant $\gamma$, and the system size $N$.

\begin{figure}
  \centering
  \includegraphics[width=0.99\columnwidth]{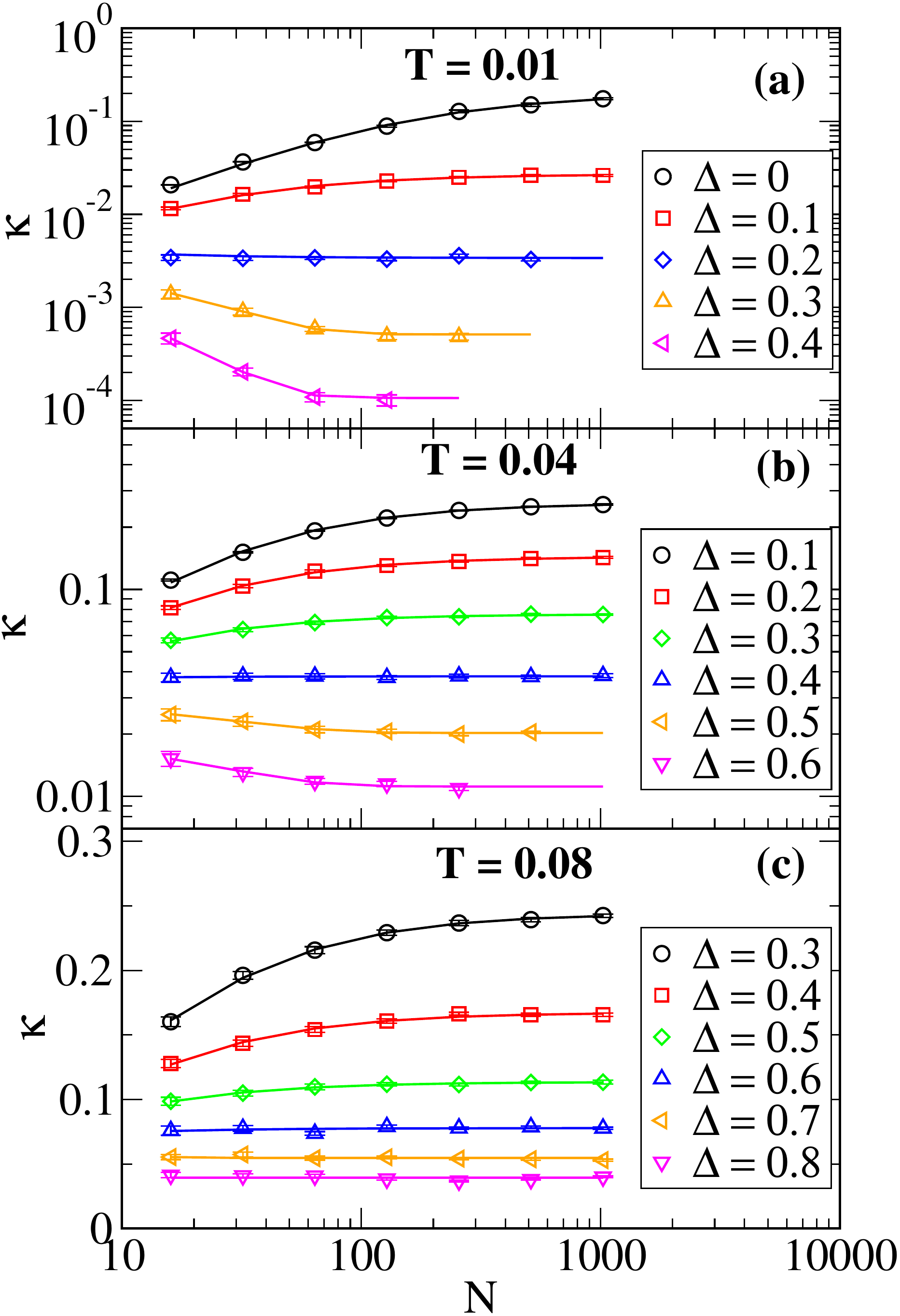}
  \caption{Plots  of $\kappa_N$ vs $N$ for different $\Delta$  for a fixed value of  $\gamma=1$,  and at different temperatures values, as specified in the frames (a), (b), and (c). Point symbols are the measured values of $\kappa_N(T)$, whereas the solid lines are plotted using Eq. \eqref{kinf_lowD} for $\Delta <\Delta_c$ and Eq. \eqref{kinf_highD} for $\Delta  > \Delta_c$.  The values of the parameters are summarized in Table.~\ref{summ_exp}. Data have been shown only for those values of $(T,~\Delta,~N)$, where a steady state profile of the local heat current $\langle J_l \rangle$ is obtained. }
  \label{KvsN_T}
  \end{figure}

\section{Simulation results for nonequilibrium steady states}
\label{sec:sim}

We compute the heat current and the temperature profile in the {\it nonequilibrium steady state} (NESS), when $T_L >T_R$.  The (scaled) heat current along the chain  from left to right is given by $J=\langle J_N \rangle= \sum_{l=2}^{N} \langle f_{l,l-1} \dot{x_l}\rangle/(N-1)$  where  $f_{l,l-1}=(x_{l-1}-x_l)^3$ is the force exerted by the $(l-1)^{\rm th}$ particle on the $l^{\rm th}$ particle.   We  define $T=(T_L+T_R)/2$. 
Then for a  finite system we define a thermal conductivity $\kappa_N(\Delta, T) = JN/(T_L-T_R)$. For a diffusive system one expects a finite value for $\kappa_\infty(\Delta,T)= \lim_{N\to \infty} \kappa_N(\Delta,T)$.
In all our numerical studies we set $(T_L-T_R)/T=0.5$ (which implies $T_L=1.25T$ and $T_R=0.75T$) and explore the system properties as we vary $\Delta,T$ and $ N$.

\begin{table}
\centering
\begin{tabular}{| c | c || c | c|  c|c|}
 \hline 
   $T$& $\Delta$ & $r$ & $\xi$ & $A$ & $\kappa_{\infty}$\\ \hline\hline
  0.01& 0   &$762\pm10$&-&-&0.200(5) \\ \cline{2-6}
  &0.1&$793\pm 36$&-&-&0.0270(4) \\ \cline{2-6}
    &0.2&$30\pm 31$&-&-&0.00340(7) \\ \cline{2-6}
    &0.3&-&$19\pm 4$&0.0020(6)&0.00052(2) \\ \cline{2-6}
        &0.4&-&$12.2\pm 2.5$&0.0014(5)&0.00010(1) \\ \hline
  0.04& 0.1   &$90\pm2$&-&-&0.263(1) \\ \cline{2-6}
      &0.2&$84.6\pm 3.5$&-&-&0.1440(9) \\ \cline{2-6}
    &0.3&$76.3\pm 6.6$&-&-&0.0761(4) \\ \cline{2-6}
    &0.4&$19\pm 4$&-&-&0.0380(4) \\ \cline{2-6}    
    &0.5&-&$32\pm 13$&0.008(5)&0.0204(2) \\ \cline{2-6}
    &0.6&-&$27\pm 15$&0.006(3)&0.0111(2) \\ \hline
 0.08& 0.3&$33.5\pm1.7$&-&-&0.244(1) \\ \cline{2-6}
     &0.4&$30.8\pm 2.5$&-&-&0.1678(8) \\ \cline{2-6} 
     &0.5&$21.7\pm 4.2$&-&-&0.1136(6) \\ \cline{2-6}
      &0.6&$12\pm 8$&-&-&0.0783(5) \\ \cline{2-6}
      &0.7&-&$308\pm 38$&0.003(1)&0.053(1) \\ \cline{2-6}
      &0.8&-&$55\pm 42$&0.005(3)&0.0383(4) \\ \hline
\end{tabular}
    \caption{A summary of exponents $r$, $\xi$, $A$, and $\kappa_{\infty}$ determined from the best non-linear fits of finite-size conductivities as shown in Fig. \ref{KvsN_T} with the form of Eqs. \eqref{kinf_lowD} and \eqref{kinf_highD}. The numbers in parentheses are the error estimates on the last significant figures.  All these data are for $\gamma=1$. }
\label{summ_exp}
\end{table}

We perform numerical simulations by using the velocity Verlet algorithm, adapted for Langevin dynamics \cite{allen2017computer}.  To speed up relaxation to the steady state, the initial conditions are chosen from a product form distribution corresponding to each disconnected oscillator being in equilibrium at temperature $T_i$ that varies linearly across the chain. The system is evolved up to times ranging from $2\times 10^9 - 5\times 10^9$ time steps of step size $dt=0.005$, in order to reach its NESS, and then NESS averages are obtained over another equal number of time steps [see appendix~\ref{appen}]. 
Relaxation times increase rapidly with increasing $N$, $\Delta$, and with decreasing $T$. We also average over $50$ disorder samples, and our error bars  represent sample-to-sample variations. 
For $T \lesssim 0.01$, the conductivity becomes very small ($\lesssim 10^{-4}$) and reaching a steady state becomes computationally challenging because the fluctuations become more pronounced. Therefore, for low temperatures, to reduce the impact of such fluctuations we perform $10^{11}$ times steps for a NESS and compute  $\kappa_N$ by taking an average over the  NESS measurements for another $10^{11}$  time steps, which has been possible for  $N=32$ and $64$.

\begin{figure}
  \centering
  \includegraphics[width=0.96\columnwidth]{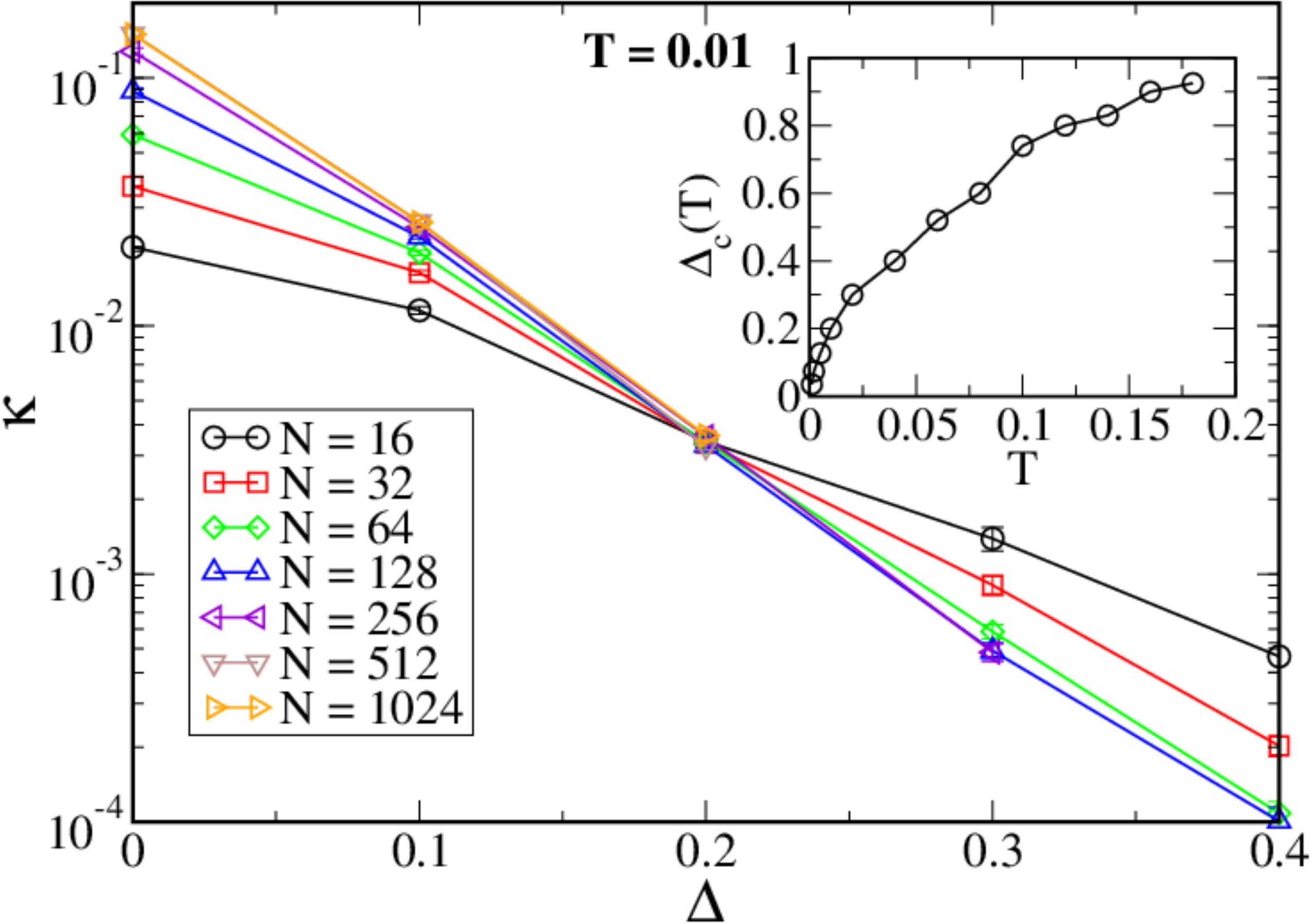}
  \caption{A plot  of $\kappa_N$ vs $\Delta$ on a log-linear scale for different $N$ at a fixed $T=0.01$ and $\gamma = 1$. An inset shows the behavior of $\Delta_c(T)$ vs $T$.}
  \label{KvsD_T0pt01}
\end{figure}

\begin{figure}
  \centering
  \includegraphics[width=0.99\columnwidth]{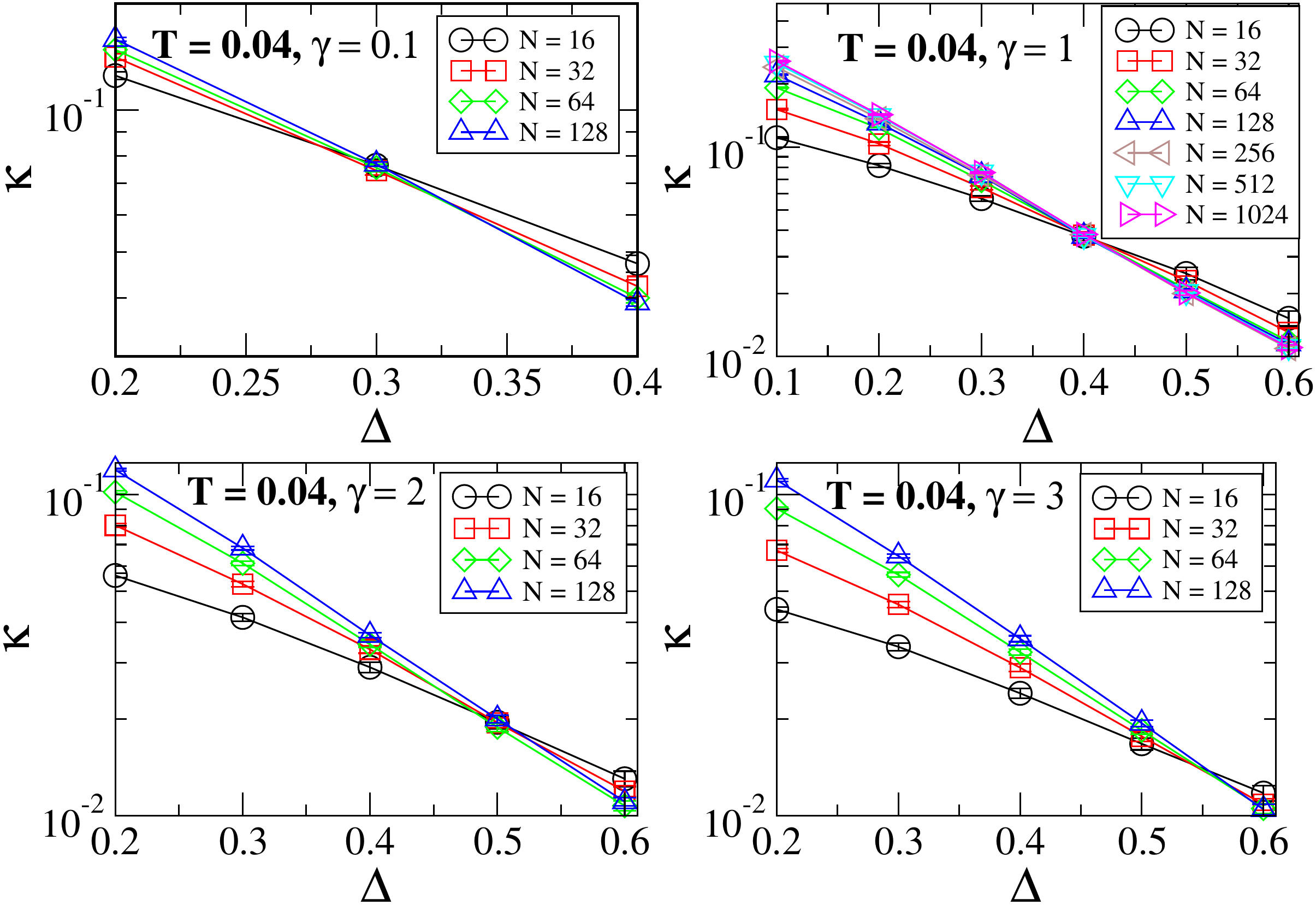}
  \caption{Plots  of $\kappa_N$ vs $\Delta$ on a log-linear scale with varying $N$ at a fix $T=0.04$, but for different values of $\gamma$. The  curves for different $N$ intersect at $\Delta_c(T=0.04,\gamma$), where $\Delta_c(T=0.04,\gamma) \simeq$ 0.3, 0.4, 0.5, 0.55 for $\gamma=$ 0.1, 1, 2, 3, respectively. }
  \label{KvsD}
  \end{figure}

In Fig.~\ref{KvsN_T} (a)-(c), we plot $\kappa_N$ against $N$ for different values of $\Delta$ at a fixed value of  $\gamma=1$, and for temperatures $T=0.01,0.04,0.08$.  We observe that in all cases, the conductivity seems to converge with increasing system size to a nonzero $\kappa_{\infty}(\Delta,T)$.  However, the approach to $\kappa_{\infty}$ with increasing $N $ is different for the small and large disorder, demarcated by a characteristic disorder strength $\Delta_c(T,\gamma)$ that depends on the temperature $T$ and the coupling $\gamma$ to the reservoirs at the ends of the chain.  For $\Delta <\Delta_c(T,\gamma)$, we find that $\kappa_N$ is an increasing function of $N$, while for 
$\Delta > \Delta_c(T,\gamma)$, it is a decreasing function.  At $\Delta= \Delta_c(T,\gamma)$ we find that the conductivity is almost independent of system size. This is illustrated in  Fig.~\ref{KvsD_T0pt01} which shows a plot of $\kappa_N$ vs $\Delta$ for different  $N$ at  $T=0.01$ and $\gamma=1$, and where the curves for different $N$ intersect at $\Delta_c \simeq 0.2$.  The variation of $\Delta_c(T,\gamma=1)$ with temperature $T$ is shown as an inset of Fig.~\ref{KvsD_T0pt01}. Fig.~\ref{KvsD} shows plots of $\kappa_N$ vs $\Delta$ for different  $N$ at a  fix $T=0.04$, but for different $\gamma$-values. Clearly, the $\Delta_c(T,\gamma)$ for a fixed $T$ increase with an increase in $\gamma$. In the following, we present all the numerical results for a fix $\gamma=1$.

\begin{figure}
  \centering
  \includegraphics[width=0.96\columnwidth]{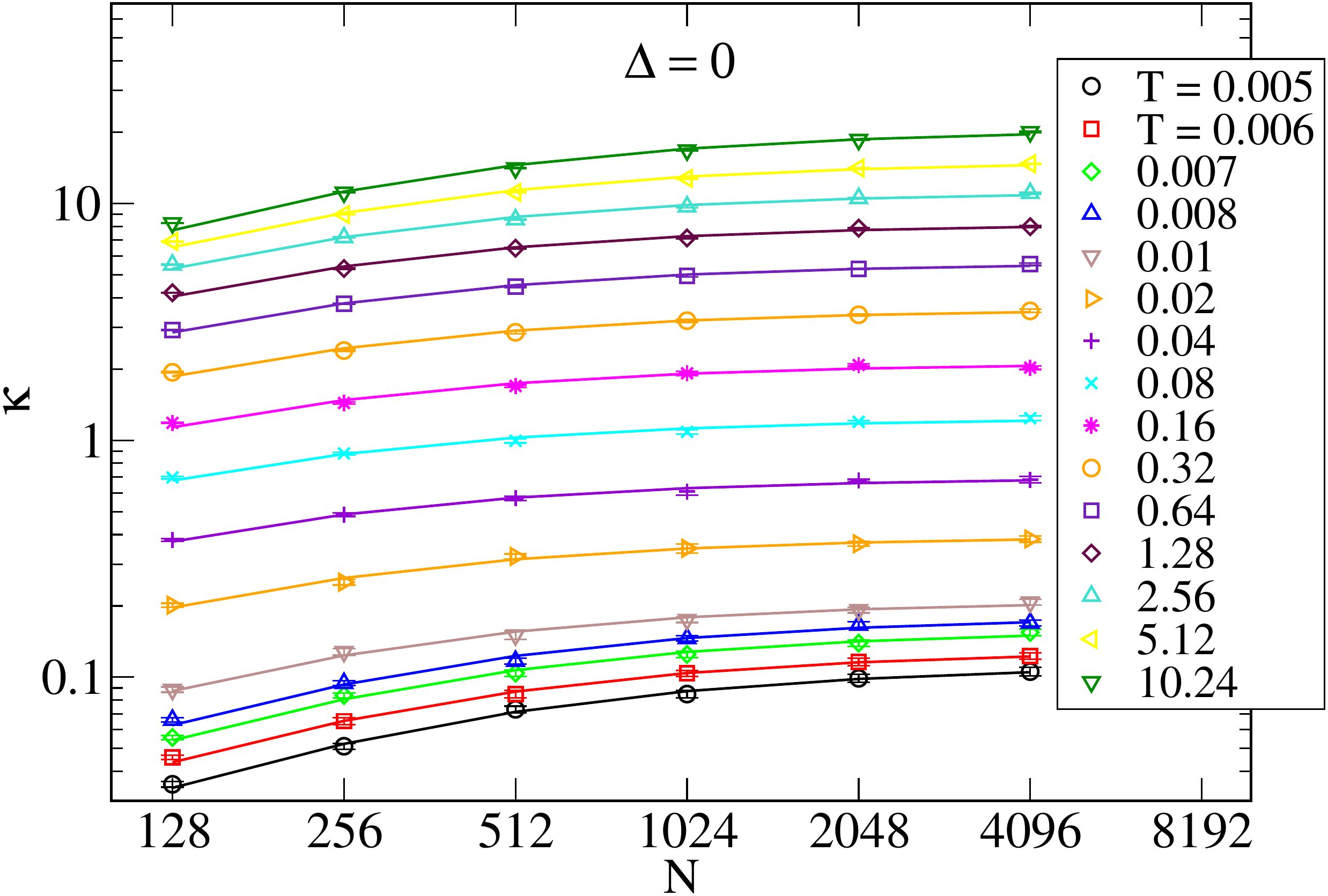}
  \caption{A plot  of $\kappa$ vs $N$ on a log-log scale for $\Delta=0$. Point symbols are the simulated values of $\kappa$ whereas the solid lines are the best nonlinear fits  of Eq. \eqref{kinf_lowD}.}
  \label{KvsN_d0}
  \end{figure}

We now discuss the difference in the system-size dependence of $\kappa_N$ between weak and strong disorder.  For weak disorder $\Delta<\Delta_c(T,\gamma)$, the bulk of the chain has a relatively low thermal resistivity $1/\kappa$, and the boundary resistance to the reservoirs at each end is high enough to produce an increase in the apparent resistivity $1/\kappa_N$ of finite length chains.
If we assume a total boundary thermal resistance $r$ due to the two ends, then the bulk and boundary resistances add to give total thermal resistance $R=(N/\kappa_\infty)+r$.  Hence the effective finite-size conductivity is given by 
 \begin{equation}
  {\kappa_N}(\Delta,T)=N/R=
 \frac{\kappa_\infty}{1+(\kappa_\infty r/N)},~ \text{for}~~ \Delta < \Delta_c.
  \label{kinf_lowD}
 \end{equation}
For  system sizes $N \ll r \kappa_\infty$ the boundary resistance dominates and one has $\kappa_N \sim N$, while for larger $N \gg r \kappa_{\infty}$ the heat transport is diffusive with $\kappa_N \sim N^0$.

\begin{figure}
  \centering
   \includegraphics[width=0.99\columnwidth]{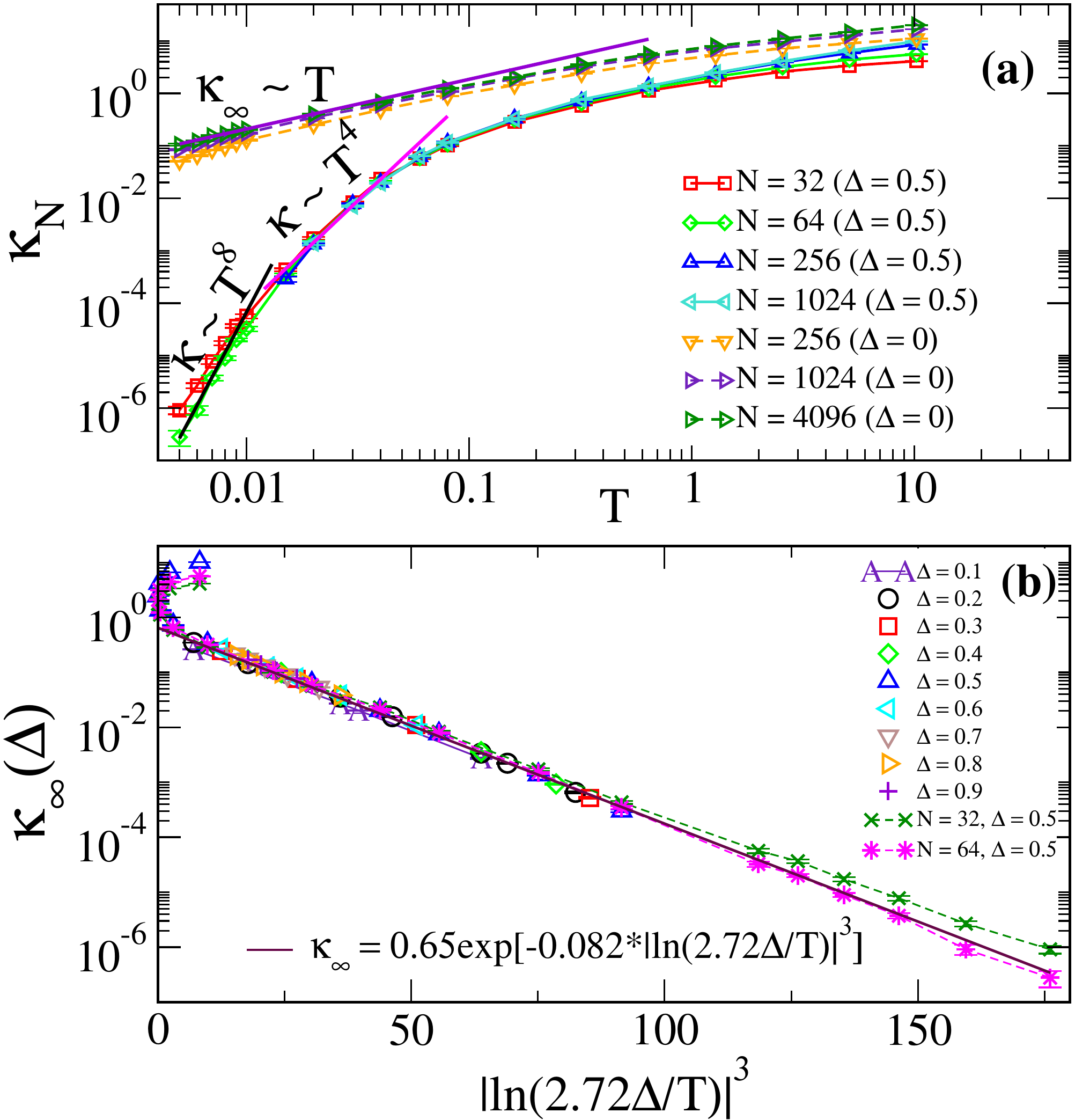}
  \caption{(a) The plot of $\kappa_N$ {\it vs.} $T$ for $\Delta=0.5$ and for the ordered case $\Delta=0$.   For the disordered case $\Delta=0.5$ we see that the slope on this log-log plot keeps increasing with decreasing $T$, and at a low $T$ with increasing $N$.  We find $\kappa \sim T^8$ in our lowest attained temperature range for $N=64$.   For the ordered case, a dependence $\kappa_\infty \sim T$ is seen at low $T$.  (b) The plot of $\kappa_\infty$  as a function of $\Delta/T$ shows a good collapse. The solid line is the fit to the form $\kappa_\infty= A \exp(-B |\ln(C \Delta/T)|^3)$. Some finite-size $\kappa_N$ data at $\Delta=0.5$ are also shown.}
 \label{KvsT_D}
\end{figure}

For the zero disorder ($\Delta =0$) case also, the finite-size effects are well described by Eq.~(\ref{kinf_lowD}) and 
Fig.~\ref{KvsN_d0} shows a plot of finite size conductivities as a function of system size $N$ with varying temperatures. For low system sizes, the heat transport is ballistic, i.e., $\kappa \sim N$, while with an increase of system size, the anharmonic oscillator potential part in the FSW model leads to a saturation of the conductivity $\kappa$. The point symbols in Fig.~\ref{KvsN_d0} are the simulated values of $\kappa$ whereas the solid lines are the best nonlinear fits  of Eq. \eqref{kinf_lowD}. Clearly,  Eq.~\eqref{kinf_lowD} fits accurately to the simulated data at all temperatures.   From these fits, the saturated  values of conductivity,  $\kappa_\infty$, can be extracted and plotted as a function of temperature $T$. We find the dependence $\kappa_\infty(T) \sim T$ at low temperatures [see Fig.~\ref{KvsT_D}(a)].

Coming to the disordered case, for strong disorder $\Delta>\Delta_c(T,\gamma)$ and low enough temperature, the short-distance, short-time behavior of the chain is insulating (Anderson localized), with the thermal conduction due to chaos being relatively weak.  This situation can be viewed as two
parallel channels of conduction:  One channel is linear conduction through Anderson-localized modes of the linearized system.  These modes couple to both reservoirs for the finite system.  Since such states decay with distance $d$ as $e^{-d/\xi(\Delta,T)}$, where $\xi$ is a localization length, their contribution to the current  $\sim e^{-N/\xi(\Delta,T)}$.  The second channel is the conduction of energy between locally chaotic multi-oscillator nonlinear resonances via the process of Arnold diffusion ~\cite{chirikov1979universal,chirikov1993theory,Basko2011}.  This leads to a small conductivity (system-size independent), which essentially gives $\kappa_\infty$. Hence the contribution from these two parallel processes suggests the following net conductivity for the finite system:
\begin{equation}
 \kappa_N(\Delta,T)=ANe^{-N/\xi}+\kappa_\infty(\Delta,T), ~ \text{for}~ \Delta>\Delta_c.
 \label{kinf_highD}
\end{equation}

As  shown in  Figs.~(\ref{KvsN_T}), the  forms in Eqs.~(\ref{kinf_lowD},\ref{kinf_highD}) provide excellent  fits (shown by solid lines) to the finite-size simulation results (plotted as point symbols) in the two different regimes (also see an appendix~\ref{appen2}). One of the fitting parameters gives the true thermal conductivity $\kappa_\infty(\Delta,T)$.  The parameters $r,~\xi,~A,$ and $\kappa_{\infty}$, obtained from our best nonlinear fits for the data of Figs.~(\ref{KvsN_T}), are tabulated in  table~\ref{summ_exp}. In this way, we fit $\kappa_N(\Delta,T)$ to the scaling forms [Eqs.~(\ref{kinf_lowD},\ref{kinf_highD}], and obtain 
$\kappa_\infty(\Delta,T)$ for many different sets of $(\Delta,T)$.

{\bf Temperature dependence of $\kappa_\infty$}:
We next study the temperature dependence of $\kappa_\infty$, particularly at low $T$.
In Fig.~\ref{KvsT_D}(a) we plot $\kappa_N(T)$ vs $T$  for $\Delta=0.5$ as well as for the pure case with $\Delta=0$.  
In both cases, the conductivity decreases with decreasing temperature and vanishes at $T=0$. 
As mentioned earlier, for the ordered case,
$\kappa_\infty \sim T$ at low $T$, while for the disordered case $\kappa_{\infty}(T)$ appears to decrease at low $T$ faster than any power of $T$.  If we fit the behavior to a power law, $\kappa_{\infty} \sim T^a$, over a narrow range of $T$, then
around $T \cong 0.02$ the effective exponent is $a \cong 4$, as was also reported in Ref.~\cite{flach2011thermal}.  Going down to $T=0.005$ and $N=64$ we find a rapid increase of this effective exponent to $a \cong 8$, which indicates that $a$ might be even larger at this $T$ for larger $N$.  In Ref.~\cite{Basko2011} it has been argued that the behavior at small $T/\Delta$ should be of $\kappa_\infty \sim e^{-B |\ln(C \Delta/T)|^3}$, with constants $B$ and $C$; we show in
Fig.~\ref{KvsT_D}(b) that the data fit rather well to this form.

\begin{figure}
  \centering
   \includegraphics[width=0.99\columnwidth]{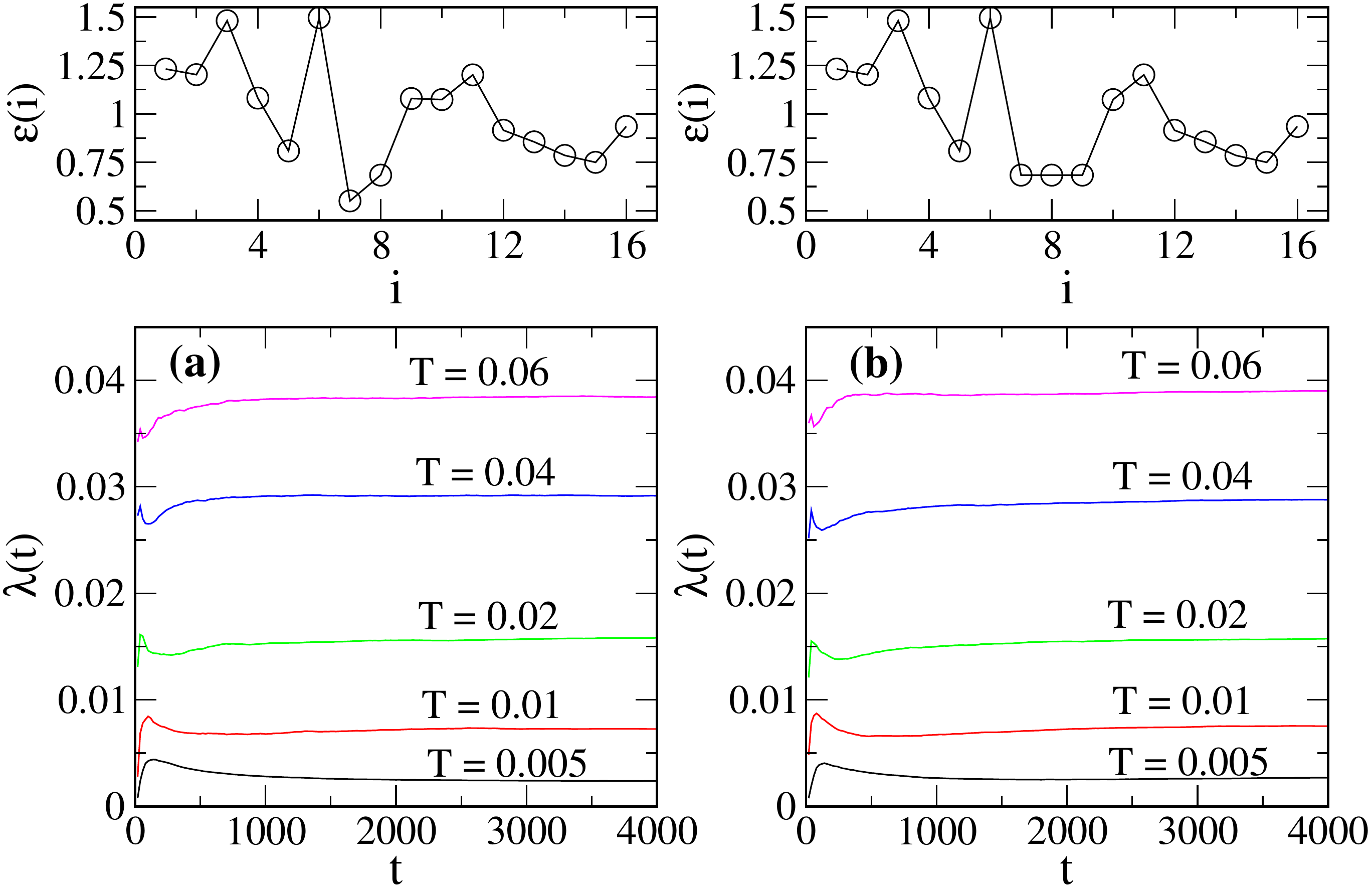}
  \caption{Finite-time Lyapunov exponents as a function of time for a single disorder sample $\{\varepsilon_i\}$ (shown on the top panel) for the case (a) [the left panel] where there are no resonances and the case (b) [the right panel]  where a three-oscillator resonance is inserted in the middle of the chain by setting $\varepsilon_{N/2\pm1}= \varepsilon_{N/2}$. In both cases, data correspond to $N=16$ with $\Delta = 0.5$ and are plotted for several temperatures as  shown in the panels.}
 \label{fig:lyap}
\end{figure}

\begin{figure}
  \centering
   \includegraphics[width=.99\columnwidth]{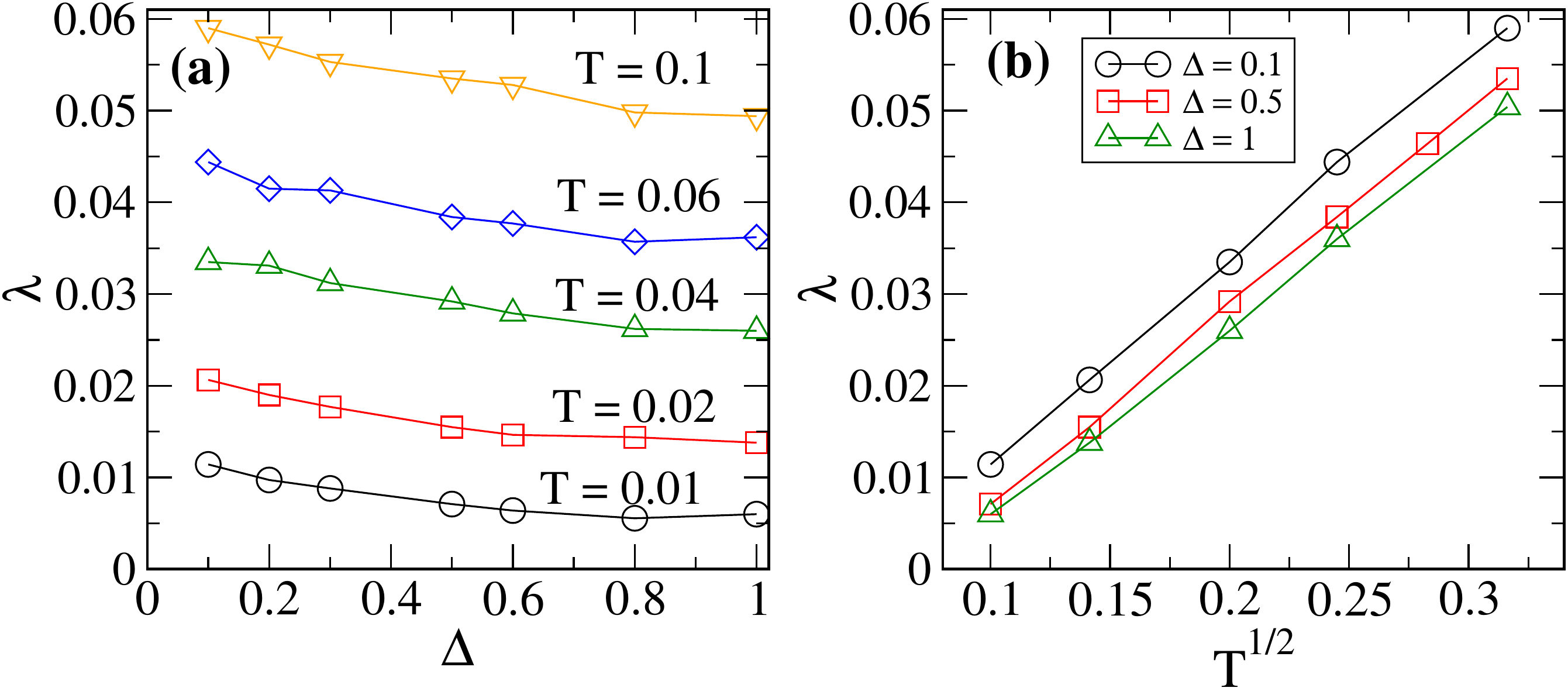}
  \caption{  (a)  Lyapunov exponent as a function of disorder strength for various temperatures. (b) The Lyapunov exponent as a function of temperature for different disorder strengths. For a fixed disorder strength, our numerics indicates that $\lambda \propto \sqrt{T}$. Here $N=16$. }
 \label{fig:lyapdelta}
\end{figure}

{\bf Transport mechanism}: There are several possibilities for the  precise mechanism by which transport occurs at low temperatures in the FSW model. One argument is based on the formation of localized chaotic islands (CI), which could provide an effective channel for energy transport. It has been argued earlier \cite{Basko2012} that the formation of such CIs requires three consecutive oscillators with resonant frequencies ($|\varepsilon_{i+1}-\varepsilon_i| \sim T$) and thus occurs with probability $p \sim T^2/\Delta^2$.  From our numerical studies with three oscillators, we found, however, that if any neighboring pair out of the three oscillators is in resonance,  this seems sufficient to generate chaos \cite{dhar2019}.  This would imply $p \sim T/\Delta$.  Since the CIs form with probability $\sim T/\Delta$, they are separated on average by distance $d \sim \Delta/T$. They would then act as effective thermal reservoirs between which energy is transmitted via intermediate localized states.  However, a detailed calculation along the lines in \cite{de2020subdiffusion} shows that one ends up with regions of large resistance and eventually sub-diffusive transport.

An alternate mechanism suggested in \cite{Basko2011} for the disordered nonlinear Schrodinger equation is that a more efficient mechanism of chaos generation does not require nearby pairs to be close in frequencies. Instead, it is possible for $n \sim \ln (\Delta/T)$ oscillators to satisfy a resonance condition and be driven to chaos by nearby sets of oscillators. Based on this picture it is estimated in \cite{Basko2011} that the probability of chaos generation scales as $  e^{-B |\ln(C \Delta/T)|^3} $ and then one can argue that the conductivity follows the same scaling.  In fact, in Fig.~\ref{fig:lyap}, we show two scenarios for $N=16$. One case (left panel) has all frequencies off-resonant, and the other case (right panel) has three oscillators in resonance. However, irrespective of whether there is resonance or not, we notice that the system is chaotic with almost the same value of the Lyapunov exponent. Therefore, as argued by \cite{Basko2011}, we also find that  chaos generation does not require nearby oscillator pairs to be in resonance. In fact, the Lyapunov exponent turns out to be independent of details of how the  random frequencies are chosen. In Fig.~\ref{fig:lyapdelta} we show the dependence of Lyapunov exponent on disorder strengths and temperatures. Interestingly, our numerics indicates that for a fixed disorder strength, $\lambda \propto \sqrt{T}$, which is similar to what is seen in several other very different classical systems \cite{moessner2018,ray2019}. In Sec.~\eqref{sec:otoc} we present further results on chaos propagation in this system.

 \begin{figure}
  \centering
  \includegraphics[width=0.98\columnwidth]{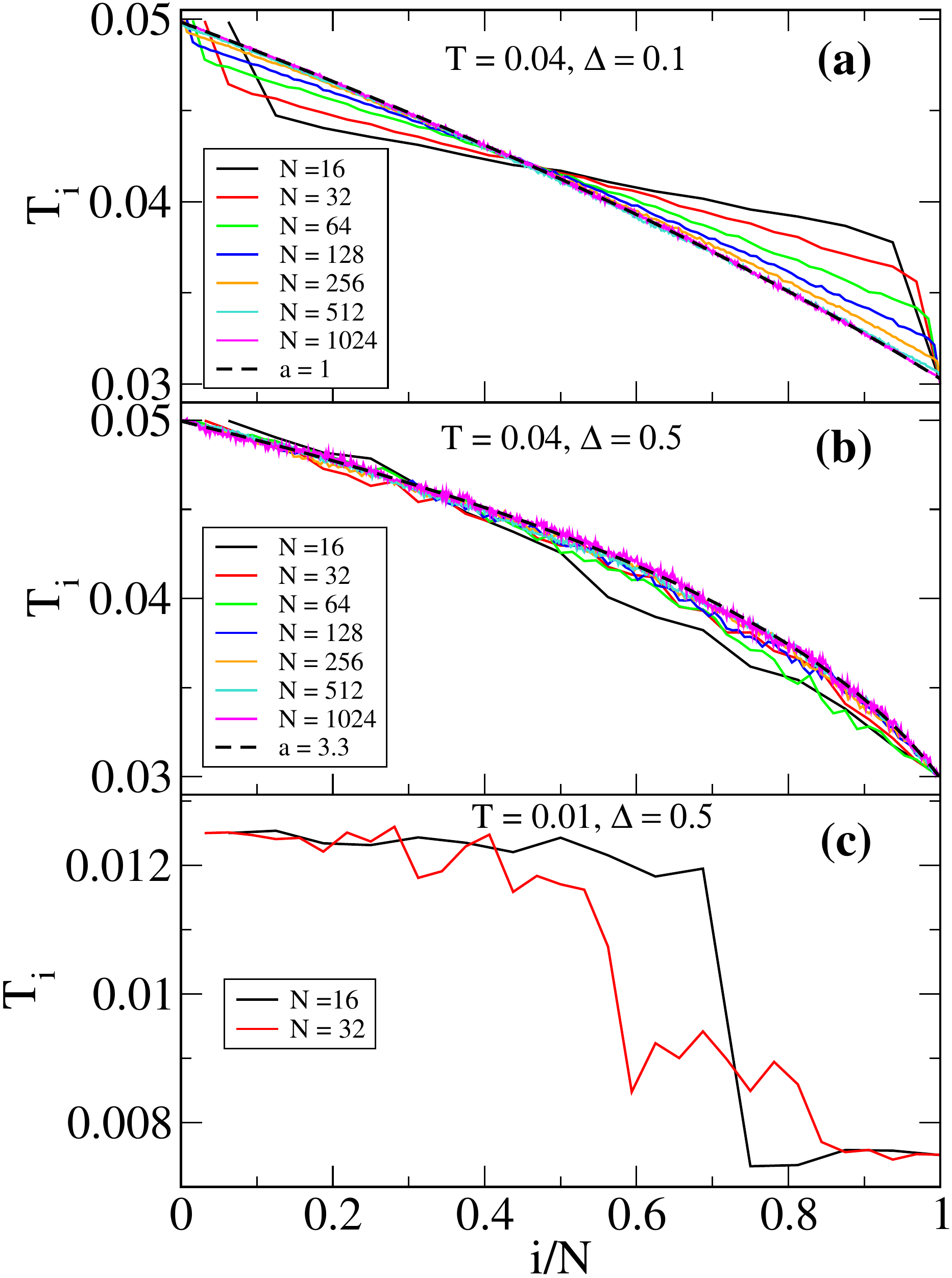}
  \caption{ (a) Temperature profiles in the steady state with $T_L=0.05$, $T_R=0.03$, and  $\Delta=0.1$ for different system sizes.
(b) Temperature profiles for $T_L,T_R$ as in (a) but with $\Delta=0.5$.  The analytical fits (black dashed lines) to the asymptotic profiles were obtained by solving $-\kappa(T)\partial_iT(i)=J$. With the form $\kappa \sim T^a$, it can be solved exactly for $T(i)$ and plotted as a function of $i/N$ using $a=1$ in (a), and $a=3.3$ in (b). (c) Temperature profile at a lower mean temperature $T=0.01$ and $\Delta=0.5$, which shows signatures  of a step profile, as seen in  MBL systems~\cite{monthus2017boundary,roeck2017}. }
 \label{tempprof}
\end{figure} 
  
{\bf Temperature profiles}:  The signatures of boundary resistance, the strong temperature dependence of $\kappa_\infty(T)$, and disorder can also be seen in the NESS temperature profiles. Note that using Fourier's law $j=-\kappa(T) dT/dx$ along with knowledge of the form $\kappa(T)$ and the boundary conditions $T_1=T_L, T_N=T_R$ uniquely fixes the temperature profile in the steady state. In  Fig.~\ref{tempprof} we plot the temperature profile $T_i= \la p_i^2 \ra$ for different temperatures and disorder strengths.
In Fig.~\ref{tempprof}(a), which is in the low-disorder regime, the boundary resistance is clearly seen for small $N$, and the profile slowly converges (with increasing $N$) to an asymptotic form that is consistent with the form $\kappa_\infty \sim T$.  
For somewhat stronger disorder and not too low temperature [Fig.~\ref{tempprof}(b)] we are near $\Delta=\Delta_c(T)$, so the profile converges quickly to the asymptotic form which is now consistent with $\kappa_\infty \sim T^{3.3}$ in this range of $T$.  These two asymptotic forms are shown by black dashed lines in Figs.~\ref{tempprof}(a) and \ref{tempprof}(b).  At even smaller temperatures and high disorder, a sufficiently small  size system is effectively in the localized regime, and we expect a step temperature profile \cite{roeck2017,monthus2017boundary}. There is some indication from our numerics that this is indeed the case [see Fig.~(\ref{tempprof}c)]. This is a signature for the classical analogue of an MBL-like regime, which, however, does not survive in the thermodynamic limit.

  \section{Simulation results for equilibrium dynamical correlation functions}
\label{sec:corr}

Equilibrium dynamical correlation functions (unequal space and unequal time) serve as another probe of transport properties and we now present results on the form of these correlations in different parameter regimes. Let us in particular focus on the spread of energy fluctuations characterized by the correlation function
\begin{align}
C(i,t)= N^{-1} \sum_{l=1}^{N} \la \left[\epsilon_{i+l}(t)-\la \epsilon_{i+l}\ra \right] [\epsilon_l(0)-\la \epsilon_{l}\ra] \ra~,
\label{corr}
\end{align}
where $\epsilon_i= p_i^2/2+k_i x_i^2/2+\nu (x_{i+1}-x_i)^4/4$ and $\la ...\ra$ denotes an equilibrium average. Here, to generate the equilibrium ensemble of initial conditions, we took the system with  periodic boundary conditions  and  attach Langevin   heat baths at temperature $T$ to every oscillator, thus ensuring  a fast equilibration.  Using initial states prepared in this way, the heat baths are then removed and the system is evolved with the Hamiltonian dynamics to compute the  time evolution of $C(i,t)$ as defined in Eq.~\ref{corr}. 

\begin{figure}
  \centering
  \includegraphics[width=0.98\columnwidth]{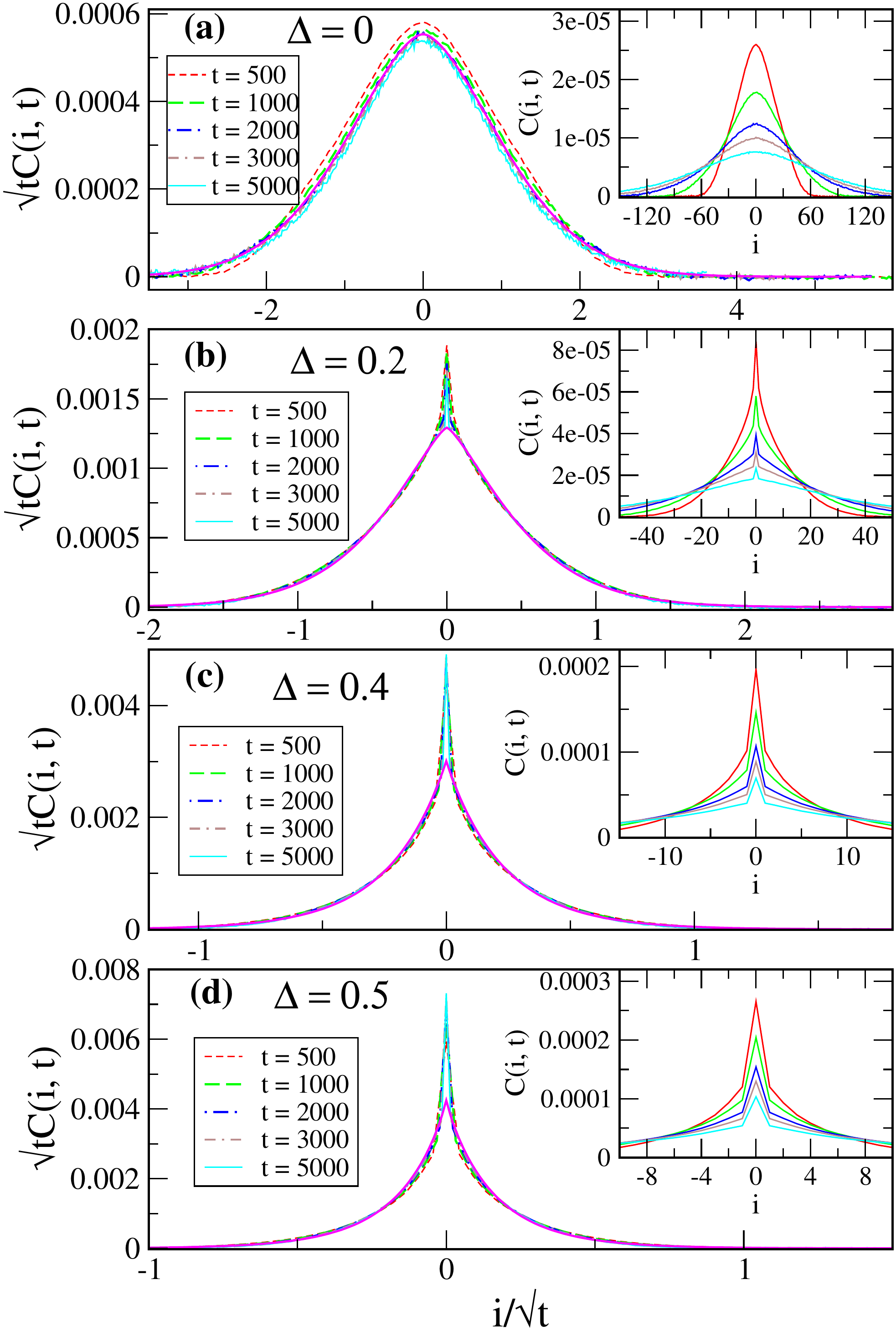}
  \caption{Diffusive scaling of energy spreading  for disorder strengths $\Delta=0,0.2,0.4,0.5$ at $T=0.04$. For this temperature $\Delta \approx 0.4$ corresponds to critical disorder. The best fit curves that are shown correspond to the form $a e^{-b |x|^c}$ with $b \approx 0.42, 2.02, 4.17, 5.43$; $c \approx 2, 1.304, 0.982,0.924$ for $\Delta=0,0.2,0.4,0.5$ respectively. The insets show the unscaled data.}
 \label{fig:corr}
\end{figure}

In Fig.~(\ref{fig:corr}) we show the time evolution of $C(i,t)$ at $T=0.04$ for four different disorder strengths. 
We find diffusive scaling of the correlations at all disorder strengths, but with non-Gaussian scaling functions except for the ordered case $\Delta=0$ (at least for the space-time $(i,t)$ scales we have reached in our numerics). A possible explanation would be that the system has a distribution of local diffusivities, which can lead to such non-Gaussian forms, yet diffusive scaling (see, for example \cite{chechkin2017,chubynsky2014diffusing}).  However, we expect that,  in the very long-time limit (inaccessible in our numerics), the scaling form will eventually become a Gaussian for the disordered case, as suggested by the observation of the absence of MBL in the previous section.  

We note  that the diffusion constant can be independently obtained using $D = \kappa_N/c_v$ where $c_v$ is the specific heat. We find the values of $D= 0.603$, 0.1528, 0.04147, 0.0223 for $\Delta=0$, 0.2, 0.4, 0.5, respectively. For the ordered case, the value of $D$ obtained in this way is consistent with the diffusion constant obtained by fitting a Gaussian in Fig.~\ref{fig:corr}(a).  Due to the fact that the diffusion constant turns out to be very small for the disordered case, therefore one needs to go to extremely long-times to see Gaussian behavior. From our studies of the equilibrium correlations we find that there is no qualitative difference in their forms between the weak and strong disorder regimes. This seems consistent with the picture that the differences that we see in the nonequilibrium studies basically arise from boundary effects. 

\section{Numerical results on chaos propagation and Out-of-time-ordered-commutator (OTOC)}
\label{sec:otoc}

Finally, we investigate the differences in chaos propagation in this system in the two regimes of weak and strong chaos. In quantum systems,  this has been studied through the so-called Out-of-Time-Ordered-Commutator (OTOC) and it is seen that chaos propagates linearly in time with a finite velocity in the conducting phase while in the MBL phase, the growth is logarithmic \cite{kim2014local,huang2017out,slagle2017out}. As the classical  analogue of the OTOC, one replaces the commutator by the Poisson bracket ($\{\cdots\}_{\rm PB}$) and this leads to  an observable \cite{das2018light} which essentially measures how an initial perturbation at the site $i=N/2$ grows in space and time. This is straightforward to compute using a linearized dynamics. 

We start with the Hamiltonian equations of motion of the system 
\begin{align}
 \dot{x_i}&=p_i,~\dot{p_i}=f_i,~~{i=1,2,\ldots,N}, {~~\rm where}  \label{hamilt_dynamic}\\
  f_i&=-\frac{\partial \mathcal H}{\partial x_i}=-k_i x_i-\nu\left[(x_i-x_{i-1})^3+(x_i-x_{i+1})^3\right].\nn
\end{align}
Let us consider an infinitesimal perturbation $\{\delta x_i(0)=0, \delta p_i(0)=\delta_{i,N/2}\}$ at site $i=N/2$ at time $t=0$ to any specific initial condition of positions and the momenta of the oscillators ($X(0)=\{x_i(0)\},~ P(0)=\{p_i(0)\}$).  Our aim is to study how  this initial localized perturbation spreads and grows through the system both in space as well as in time.  In order to do so, we look at the OTOC $D(r,t)$ defined as 
\begin{align}
D(r,t)=
\left.\{p_{(N/2+r)}(t),x_{N/2}(0)\}\right._{\rm PB}^2=\left(\frac{\partial p_{(N/2+r)}(t)}{\partial p_{N/2}(0)}\right)^2.
\label{otoc_def}
\end{align}
 From the linearized form of the equations of motion in Eq.~\eqref{hamilt_dynamic}, the  evolution of the perturbation is given by 
 \begin{align}
\dot{\delta x_i}&= \delta p_i \nn \\
\dot{\delta p_i} &= -k_i \delta x_i-3\nu \big[(x_i-x_{i-1})^2(\delta x_i-\delta x_{i-1}) \nonumber \\& ~~~~~~~~~ +~ (x_i-x_{i+1})^2(\delta x_i-\delta x_{i+1})\big],
  \label{perturb_dynamic}
 \end{align}
for $i=1,2,\ldots,N$. 

The quantity of interest for measuring the spreading of a localized perturbation  given in Eq.~\ref{otoc_def} can be rewritten as 
\begin{equation}
D(r,t)=[\delta p_{(N/2+r)}(t)]^2.
\end{equation}
 To obtain $\delta p_{(N/2+r)}(t)$, we need to solve the system of equations in \eqref{hamilt_dynamic} and \eqref{perturb_dynamic}. 
We solve these ODEs using a fourth-order Runge-Kutta (RK4) numerical integration scheme with a  time step $dt=0.005$ and with periodic boundary condition $(x_0=x_N,~x_{N+1}=x_1$). The initial condition of $X(0),~ P(0)$ is chosen from the equilibrium Gibbs distribution, $\rho(X(0), P(0))=e^{-\beta\mathcal{H}}/Z$, where $Z=\int dXdP ~ e^{-\beta\mathcal{H}}$ is the partition function. For our nonlinear model, this initial condition can easily be generated by connecting  all sites to the Langevin heat baths at the same temperature $T$. In this way, the system equilibrates very fast and the distribution of $\{x_i,p_i\}$ follow the equilibrium Gibbs distribution.

\begin{figure}
  \centering
  \includegraphics[scale=0.37]{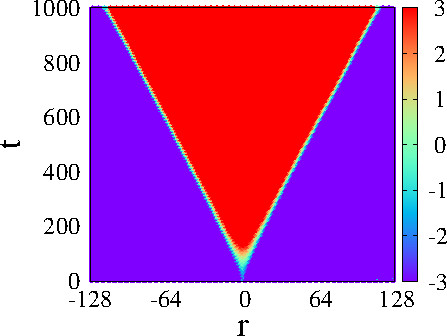}
  \includegraphics[scale=0.37]{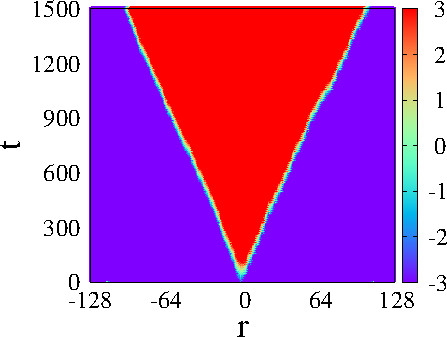}
\caption{Heat maps showing the spatio-temporal growth of the OTOC for a system of size $N=256$ in the (a) weak disorder regime ($\Delta=0.1$) [left panel] and (b) strong disorder case ($\Delta=0.6$) [right panel]. The map shows the strength of $\langle \ln D(r,t) \rangle$ where the average is over $10000$ initial configurations drawn from an equilibrium distribution at temperature $T=0.04$.}
  \label{heatmap}
\end{figure}

For a chaotic system, it is expected that the signal should arrive at the site $r$ at  a time $t_r=r/c$ where $c$ gives the speed of chaos propagation (we define the arrival time through $D(r,t_r)=1$). At long  times the signal would eventually grow exponentially with time with Lyapunov exponent $ \lambda = \langle \ln D(r,t) \rangle / 2 t$. It is to be noted that $\langle ... \rangle$ denotes the average over equilibrated initial conditions $(X(0), P(0))$ and disorder realizations. For a given disorder realization, the quantities $(c,\lambda)$ which characterize chaos propagation depend on initial conditions, and we thus study the averaged quantity  $\langle \ln D(r,t) \rangle$.  

In Figs.~(\ref{heatmap}a,b) we display heat maps showing the space-time evolution 
of $\langle \ln D(r,t) \rangle$ in the weak and strong disorder regimes respectively for a chain of size $N=256$. Unlike the quantum case, here we do not see (even at early times)  any signature of logarithmic growth in the strong disorder case.  We see ballistic propagation in both 
cases with a notable difference in the magnitude of the speed and the Lyapunov exponent. 
%

\begin{figure}
  \centering
  \includegraphics[width=0.95\columnwidth]{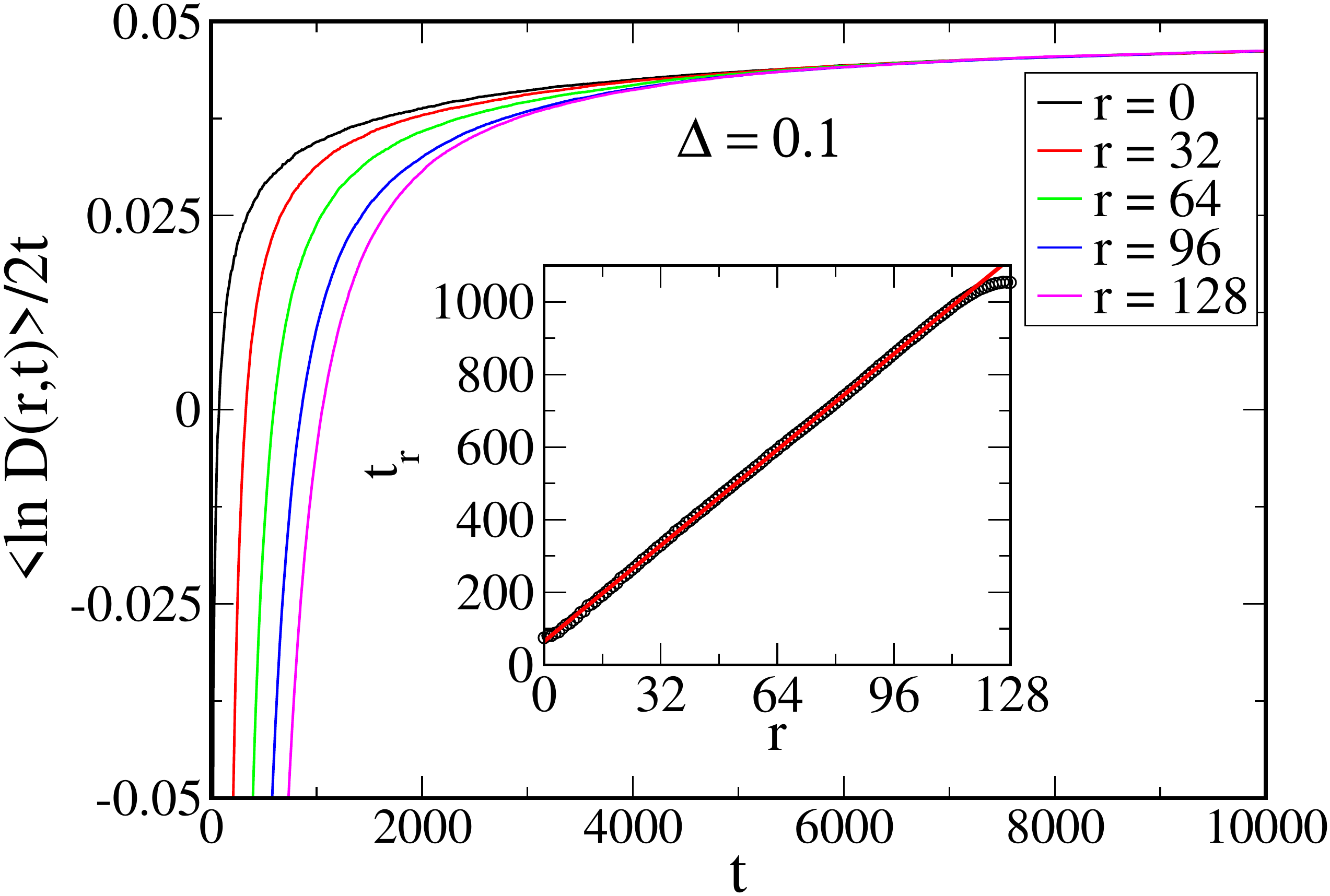}
  \caption{A plot of $\langle \ln D(r,t) \rangle /2t$ with time $t$ for $\Delta=0.1$ and at different values of $r$ . The inset shows a linear behavior of $t_r$  with $r$, for which  $D(r,t_r)=1$, and a solid red line is the best linear fit. Here $T=0.04$.}
  \label{otoc_d0pt1}
  \end{figure}

\begin{figure}
  \centering
  \includegraphics[width=0.95\columnwidth]{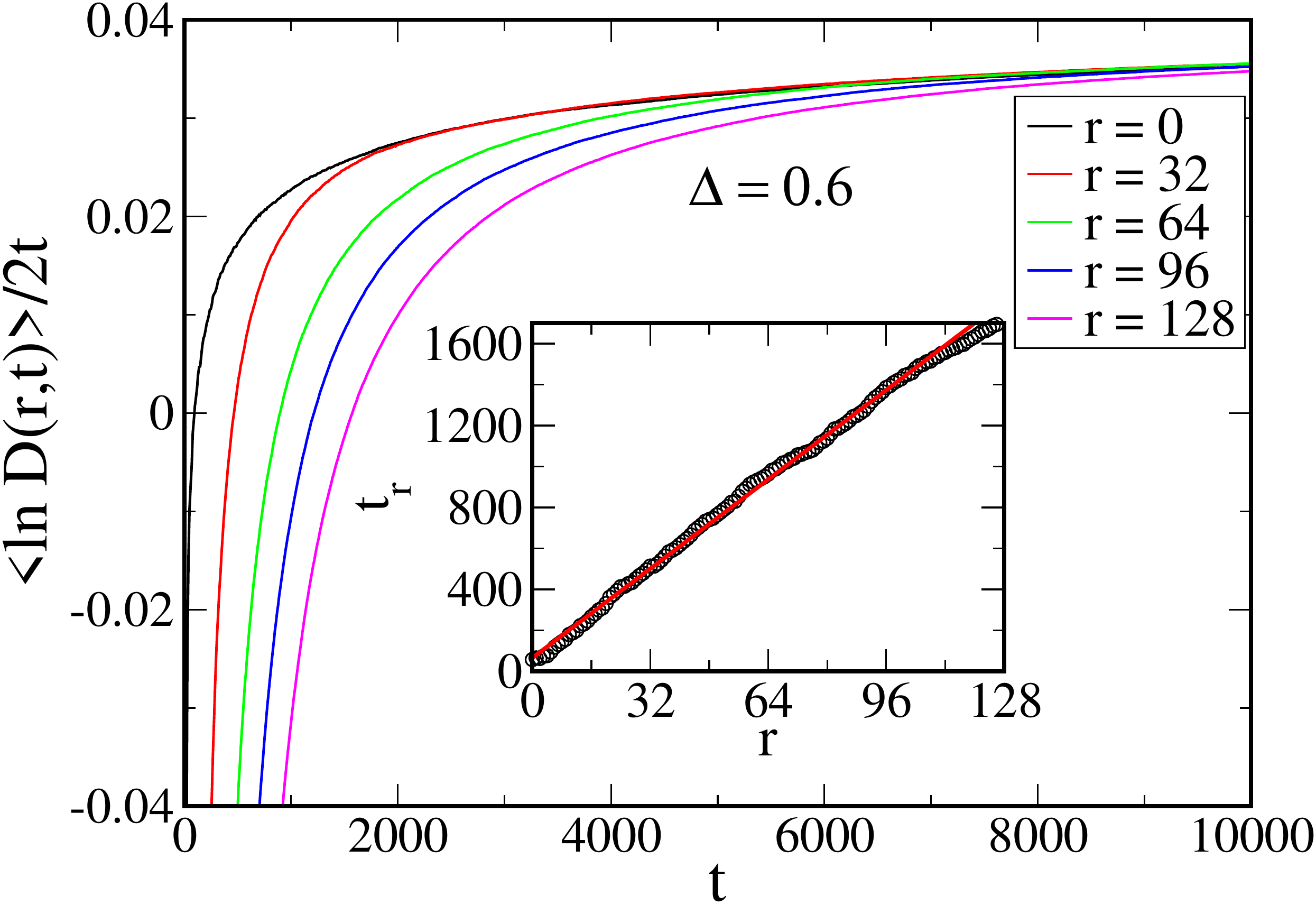}
  \caption{Similar to Fig.~\ref{otoc_d0pt1}, but for $\Delta=0.6$.}   
  \label{otoc_d0pt6}
  \end{figure}

  Fig.~\ref{otoc_d0pt1} shows a plot of $\langle \ln D(r,t) \rangle /2t$ with time $t$ at different values of $r$ for $\Delta=0.1$.  The quantity $\langle \ln D(r,t) \rangle /2t$ gives the Lyapunov exponent $\lambda$ in the limit $t\rightarrow \infty$. The numerical data are averaged over $10^4$ such equilibrated initial conditions at a temperature $T=0.04$. At large time $t$, the curves for different $r$ saturates at a value $\lambda=0.046$. Next, we define $t_r$ as a time at which the Lyapunov exponent vanishes, i.e., $D(r,t_r)=1$. The inset in Fig.~\ref{otoc_d0pt1} shows a plot of $t_r$ vs $r$, and a solid line is the best linear fit. From the slope of this fit, the speed of growth of the perturbation is given by $c=1/8.243 \simeq 0.1213$.  Fig.~\ref{otoc_d0pt6} shows a similar plot for higher disorder $\Delta =0.6$. Here we found $\lambda=0.0352$ and the speed $c=1/13.62 \simeq 0.0734$. To summarize, from our simulations we estimate $(c,\lambda)\approx(0.1213,0.046)$ for $\Delta=0.1$ and $(c,\lambda)\approx(0.0734, 0.0352)$ for $\Delta=0.6$ at $T=0.04$ and $N=256$. We see that as one increases disorder strength, both the butterfly velocity and Lyapunov exponent decrease. Note that the Lyapunov exponents are somewhat larger than the ones reported in Fig.~\eqref{fig:lyapdelta} in Sec.~\eqref{sec:sim}. This is because of the smaller system size studied there ($N=16$).

\section{Conclusions and Outlook}
\label{conclu}

 We studied the transport properties of a nonlinear chain with the weak and strong disorder, and looked for signatures of classical many-body localization.  From our  numerical studies of the nonequilibrium steady state, we find an interesting cross-over behavior whereby  the system-size scaling of conductivity ($\kappa_N$) is qualitatively different above and below  a characteristic disorder ($\Delta_c$), that depends not only on temperature but also on the  coupling strength to the baths.  
We find that the finite-size effects in the thermal conductivity are consistent with boundary effects.  On the one hand, there is a regime of weak enough disorder where the system can be viewed as thermal resistors in series, one for the length of the oscillator chain and the others for the couplings between the ends of the chain and the heat reservoirs.  In this regime the boundary resistances suppress the measured thermal conductance.  On the other hand, at low enough temperature the nonlinearity and thus the chaos are weak, and for strong enough disorder the system can be approximately realized as linearized, resulting in Anderson localized modes.  In this regime short chains can be viewed as having two parallel channels for thermal conduction, one directly from reservoir to reservoir via the localized modes of the chain and the other through the weakly chaotic diffusive transport within the bulk of the chain.  In this regime the extra conductance via the localized modes enhances the measured thermal conductance of short chains.

Our finite-size scaling analysis leads to estimates of the thermodynamic limit conductivity $\kappa_\infty$ and we find evidence that for strong disorder $\kappa_\infty$ is a function of the scaled variable $\Delta/T$. Our data  are described well by the form  $\kappa_\infty \sim  e^{-B |\ln(C \Delta/T)|^3} $ which is consistent with Ref.~\cite{Basko2011} for the discrete nonlinear Schrodinger equation.  As argued in this reference, our numerical studies also suggest that chaos results from many-particle resonances rather than a  few particle ones.
The form of $\kappa$ is quite non-trivial and a similar form for time-scales associated with the spreading of perturbations in disordered nonlinear media was  suggested earlier in \cite{benettin1988}.
We also investigated the temperature profile in NESS, and for strong disorder and low temperatures, we found hints of step-like profiles, a feature that is expected in systems with localization \cite{monthus2017boundary,roeck2017}. 

We do not see signatures of the weak-strong chaos cross-over in the form of equilibrium correlation functions which exhibit diffusive scaling in both the weak and strong disorder regimes, as expected since the cross-over is dominated by boundary effects.  We  find strong non-Gaussianity which we expect would go away in the long time limit.  A study of the OTOC in the two regimes shows that chaos propagation is always ballistic though the butterfly speed and the Lyapunov exponent are  smaller in the strong disorder regime. We observed a $T^{1/2}$ dependence of the Lyapunov exponent on temperature for both strong and weak disorder.

Our study naturally leads to asking similar questions (such as spread of the perturbations) in models in which the oscillators in space are coupled even in the absence of nonlinearity. This could provide more insight into many-body localization transition in classical systems. Future work also includes understanding transport and OTOC in a model where disorder has a fractal pattern \cite{Narayan2018} which has been proposed as the closest classical analogue to many-body localization. Needless to mention, a rigorous understanding of the transport mechanism at ultra-low temperatures in a nonlinear disordered many-body classical system remains an interesting open problem.

\acknowledgements

 We thank F. Huveneers and W. De Roeck for many useful discussions and for pointing out errors in an earlier analysis. We also thank C. Dasgupta for useful discussions. Manoj Kumar would like to acknowledge  ICTS postdoctoral fellowship and the Royal Society - SERB Newton International   fellowship (NIF$\backslash$R1$\backslash$180386). AK acknowledges support from DST grant under project No. ECR/2017/000634.
 MK gratefully acknowledges the Ramanujan Fellowship SB/S2/RJN-114/2016 from the Science and Engineering Research Board (SERB), Department of Science and Technology, Government of India.  
 AD and AK would like to acknowledge support from the project 5604-2 of the Indo-French Centre for the Promotion of Advanced Research (IFCPAR).
  AD, AK, and MK acknowledge support of the Department of Atomic Energy, Government of India, under project no.12-R\&D-TFR-5.10-1100 and would also like to acknowledge the ICTS program on ``Thermalization, Many body localization and Hydrodynamics (Code: ICTS/hydrodynamics2019/11)'' for enabling crucial discussions related to this work.  
    DH is supported in part by a Simons Fellowship and by (USA) DOE grant DE-SC0016244. The numerical simulations were performed on a {\it Mario} HPC at ICTS-TIFR and a {\it Zeus} HPC of Coventry University.


\appendix
\section{Nonequilibrium steady state of the heat current}

  \label{appen}

  \begin{figure}
  \centering
  \includegraphics[width=0.99\columnwidth]{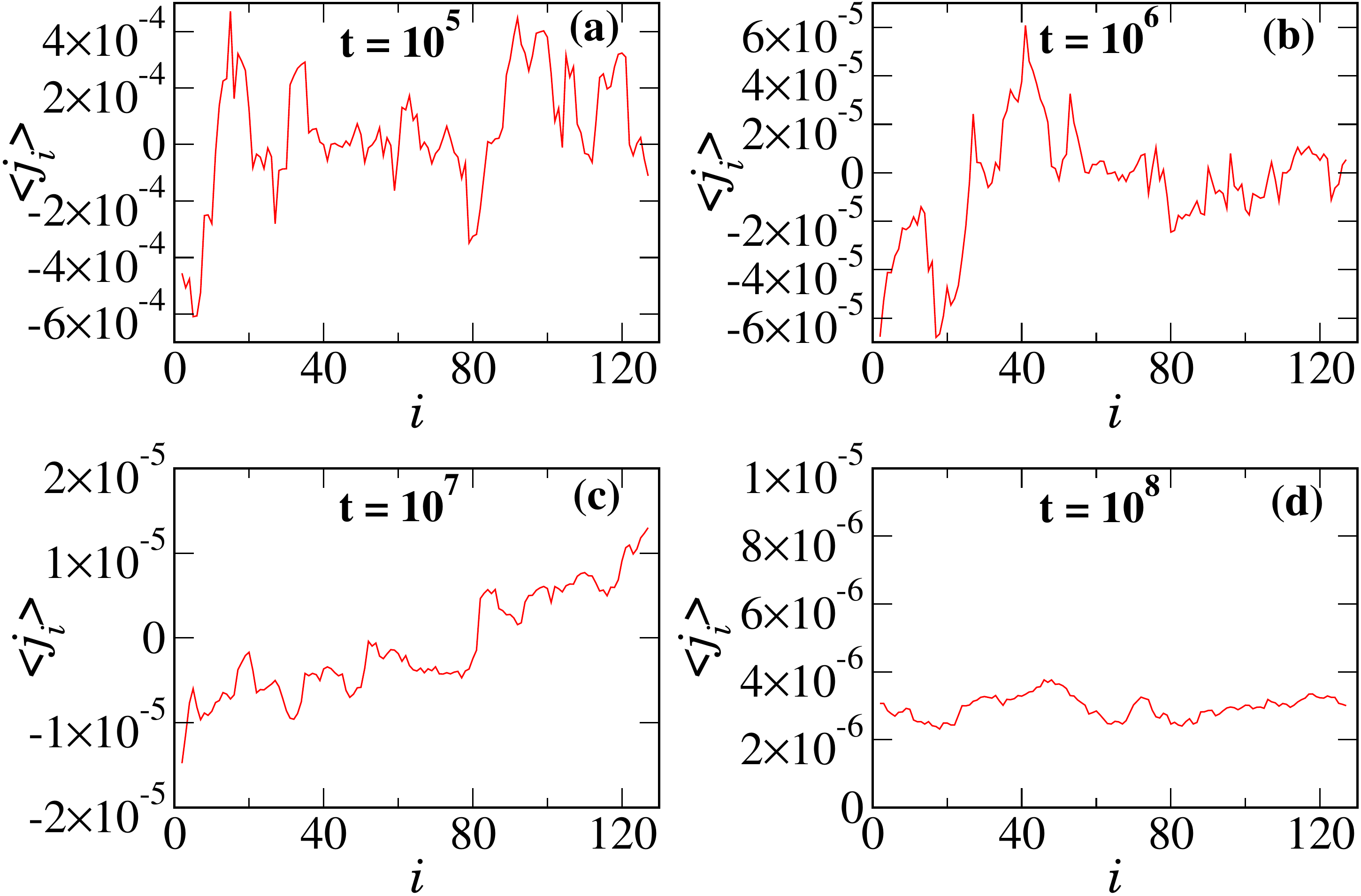}
  \caption{Energy current profiles, $\langle j_i \rangle$ versus
 $i$, of the system of size $N=128$  at $\Delta=0.5$, and  $T=0.04$.  We plot these profiles for different amounts of time $t$ as specified in the panels (a) to (d).}
  \label{curr_Tpt04}
  \end{figure}

  \begin{figure}
  \centering
  \includegraphics[width=0.99\columnwidth]{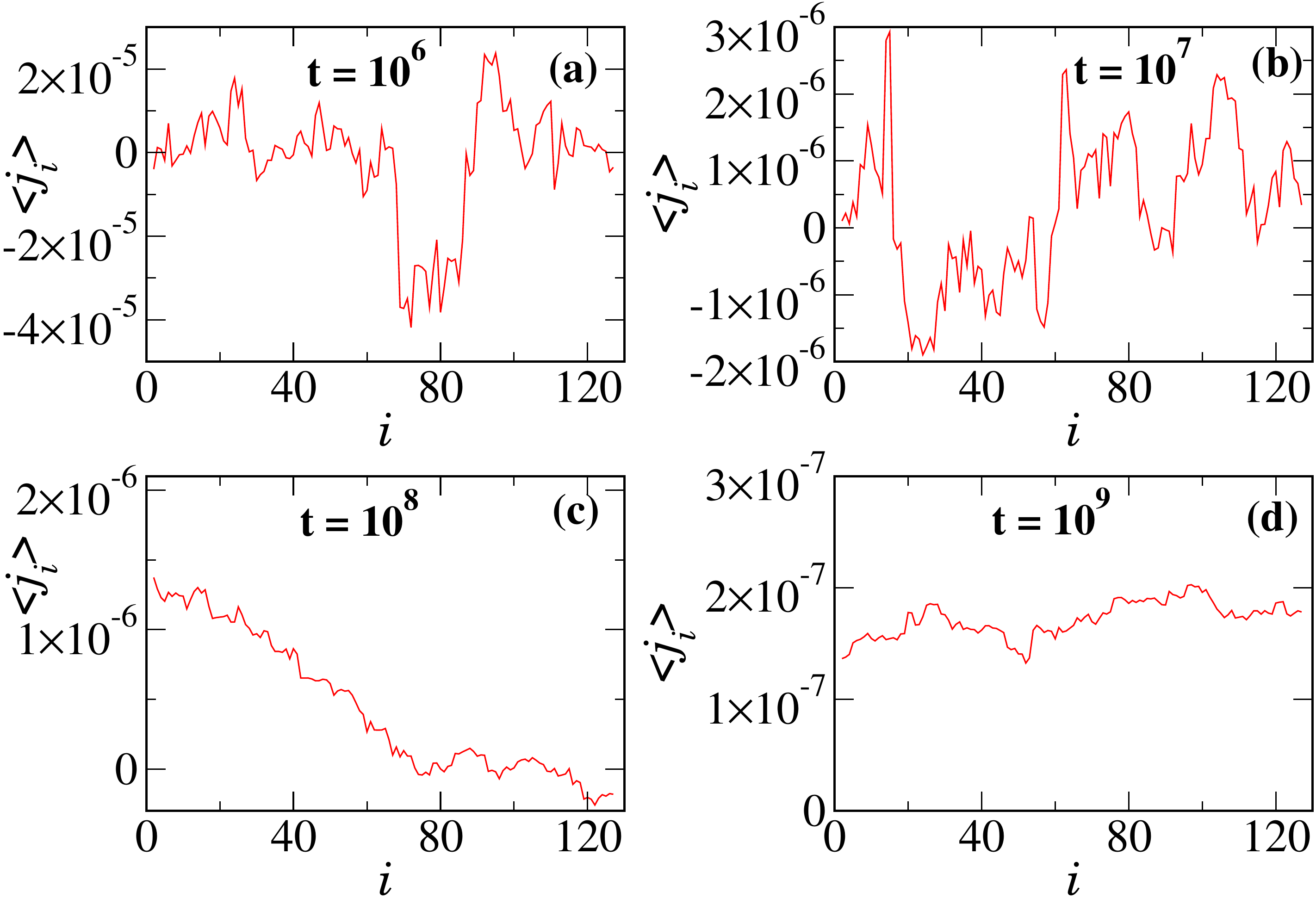}
  \caption{Analogous to Fig.~\ref{curr_Tpt04}, but for $T=0.02$. }
  \label{curr_Tpt02}
  \end{figure}

For our one-dimensional system of $N$ oscillators connected to two heat baths at its end, the time derivative of energy $\epsilon_l$ associated with $l^{\rm th}$ particle or oscillator, in terms of current $j_{l,l-1}$ from $l-1$ site to $l$, is given as \cite{dhar2008heat}  
\begin{eqnarray}
\label{ene_l}
\dot{\epsilon_1}   &=& -j_{2,1}+j_{1,L},  \nonumber\\
\dot{\epsilon_l}&=& -j_{l+1,l}+j_{l,l-1}\quad \text {for} ~~ l=2,3,\cdots,N-1,  \nonumber\\
\dot{\epsilon_N}   &=& j_{N,R}+j_{N,N-1},
\end{eqnarray}
where $j_{l,l-1}=1/2(\dot x_{l-1}+\dot x_l)f_{l,l-1}$ with $f_{l,l-1}=-\partial U(x_{l-1}-x_{l})/\partial x_l=(x_{l-1}-x_{l})^3$. $j_{1,L}$ and $j_{N,R}$ are the instantaneous energy current from the left and right reservoirs into the system, respectively. These are given as $j_{1,L}=f_L\dot x_1=(-\gamma \dot x_1 +\eta_L)\dot x_1$ and $j_{N,R}=f_R\dot x_N=(-\gamma \dot x_N +\eta_R)\dot x_N$.

In the steady state, if we denote the time average as $\langle \cdots \rangle$, the Eq.~\eqref{ene_l} then demands the equality of current flowing between any neighboring pair of particles, i.e,
\begin{equation}
\label{ness}
   \langle j_{1,L} \rangle= \langle j_{2,1} \rangle= \langle j_{3,2} \rangle=\cdots \langle j_{N,N-1} \rangle= -\langle j_{N,R} \rangle,
\end{equation}
with $\langle j_{l,l-1} \rangle = \langle \dot x_l f_{l,l-1} \rangle$ upon using 
$\langle \dot x_{l-1} f_{l,l-1}\rangle=\langle \dot x_l f_{l,l-1}\rangle$. Thus, in order to reach the steady state of the system, we determine the energy current $\langle j_{l,l-1} \rangle$ between all neighboring pair of particles, and examine the behavior of  $\langle j_{l,l-1} \rangle$ versus $l$ for different amounts of time. A nonequilibrium steady state (NESS) is reached when Eq.~\ref{ness} holds, i.e.,  the current profile of the system, for  $\langle j_{l,l-1} \rangle$ or in simple notation $\langle j_l \rangle$ versus $l$, is showing essentially a flat behavior. 
\begin{figure}
  \centering
   \includegraphics[width=0.99\columnwidth]{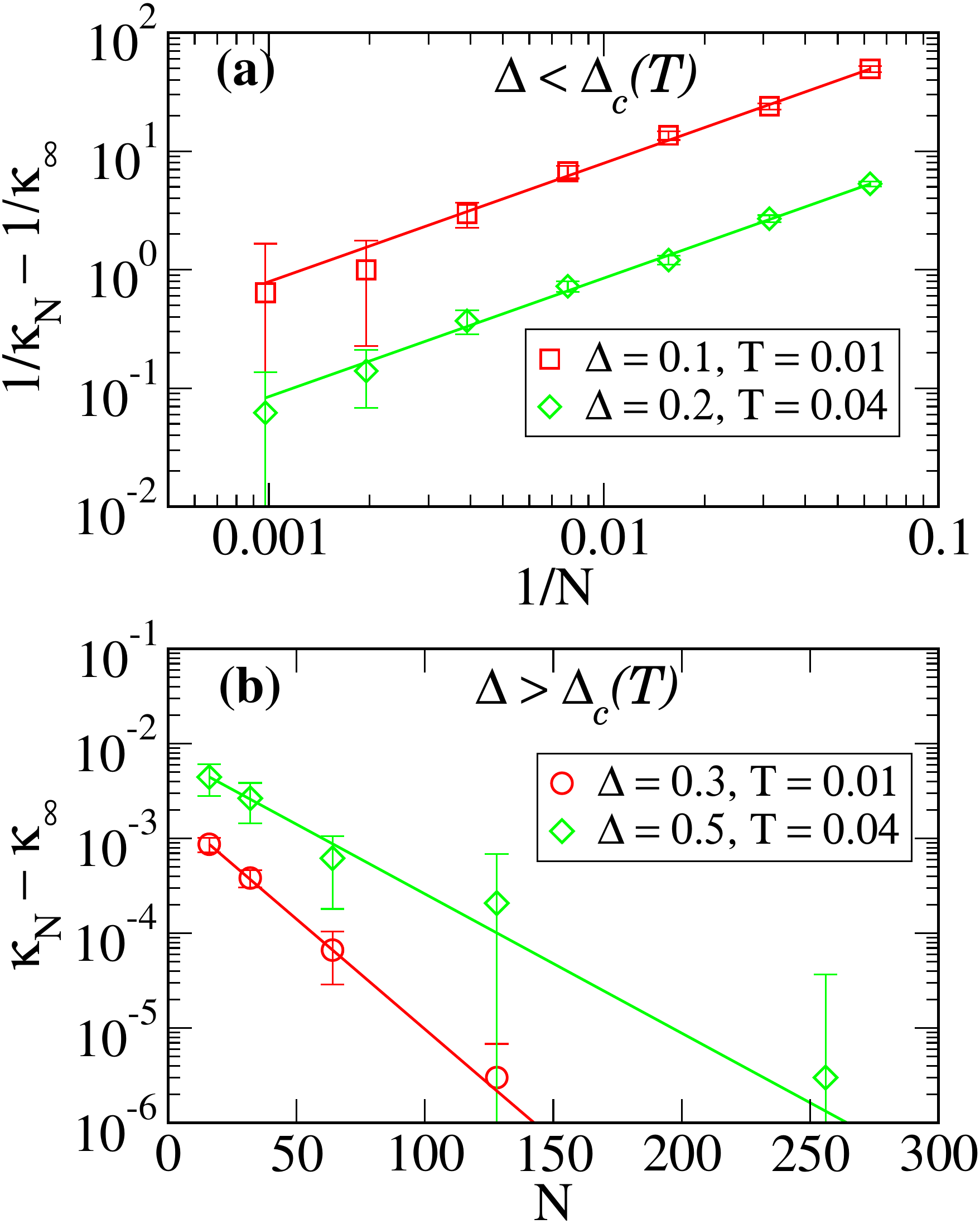}

  \caption{(a) The plot of $1/\kappa_N-1/\kappa_\infty$ versus $1/N$ (on a log-log scale) for $\Delta<\Delta_c(T)$, where $\Delta_c(T=0.01)\simeq0.2$, and $\Delta_c(T=0.04)\simeq0.4$. (b) The plot of $\kappa_N-\kappa_\infty$ versus $N$ (on a log-linear scale) for $\Delta>\Delta_c(T)$. The point symbols are the simulated values for $\kappa_N(\Delta, T)$, which have been shown only for those $N$, to which the steady state is obtained. The solid lines are the fits of two different forms, presented in Eqs.~(\ref{kinf_lowD}) and (\ref{kinf_highD}) for $\Delta <\Delta_c (T)$ and $\Delta >\Delta_c(T)$, respectively.}
  \label{KvsN_inverted}
  \end{figure}

In Fig.~\ref{curr_Tpt04}, we show  the current profiles of the system of size $N=128$, for a particular disorder sample  at $\Delta=0.5$, and at $T=0.04$. We compute these energy currents independently for various values of time, as mentioned in the panels (a) to (d). Notice from Fig.~\ref{curr_Tpt04} the scale of fluctuations in energy current, flowing between each neighboring particle, which decays as the time $t$ is raised, and eventually a steady state is reached in time $t$ of the order of $10^8$ as seen in the panel (d). In order to check the effect of changing the disorder sample on relaxation, we repeated this same analysis for different disorder realizations drawn from the same $\Delta$-value, and found that a steady state is reached in typically the same order of relaxation time $t_{\rm eq}$. Therefore we emphasize that the relaxation time does not depend upon a disorder sample. Instead, it   depends on parameters like $N$, $\Delta$, and $T$.

With lowering $T$, $t_{\rm eq}$ increases rapidly as shown in a Fig.~\ref{curr_Tpt02}, where we plotted the energy currents for the same values of $N=128$ and $\Delta=0.5$, but at $T=0.02$. See  panel (d) of this figure, which is demonstrating  the steady state in $t$ of $O(10^9)$. Comparing Figs.~\ref{curr_Tpt04} and \ref{curr_Tpt02}, the relaxation time $t_{\rm eq}$ increases about 10 times in lowering the temperature $T=0.04$   to $T=0.02$. Thus, it is the reason that at much lower temperatures $T\lesssim 0.01$, we use $t_{\rm eq} \simeq 10^{11}$. With this procedure of reaching a steady state, we then started our measurement to compute NESS averaged heat current and also averaged it over several disorder samples.

\section{Finite-size scaling corrections of the thermal conductivity}
  \label{appen2}

To look for any dominant finite-size corrections in the scaling of conductivity given in Eqs.~(\ref{kinf_lowD},\ref{kinf_highD}), we replot some of the data of Fig~\ref{KvsN_T} in  different manners, as shown in  Fig.~\ref{KvsN_inverted}. In particular, we plotted the residual-like quantities, $1/\kappa_N-1/\kappa_\infty$ against $1/N$ for the weak disorder $\Delta<\Delta_c(T)$, and  $\kappa_N-\kappa_\infty$ against $N$ (on a semi-log scale) for the strong disorder $\Delta>\Delta_c(T)$. The point symbols denote the simulated values, whereas solid lines are the fitting lines. In panel (a), such lines are linear fits, representing $1/\kappa_N-1/\kappa_\infty \sim 1/N$, while the lines in panel (b) are exponential fits of the form $\kappa_N-\kappa_\infty \sim \exp(-N/\xi)$. Clearly, the fits in both panels agree very well to the simulated points, implying that our data do not show the presence of any finite-size scaling corrections. Hence, Eqs.~(\ref{kinf_lowD}) and (\ref{kinf_highD}), respectively,  for $\Delta <\Delta_c$ and $\Delta >\Delta_c$, precisely describe the system size scaling of conductivity $\kappa_N$.

 \bibliography{ref.bib}

\end{document}